\documentclass[12pt]{article}  
\usepackage{amssymb,amsmath}  
\usepackage{graphicx}

\def\sst{\scriptscriptstyle}

\def\sl{\slshape}

\def\kn#1{{\kern -#1 true cm}}

\newtheorem{thm}{Theorem}

\newtheorem{cor}{Corollary}

\newcommand{\ci}{\cite}
\def\lab{\label}
\newcommand\tlab{\label}
\newcommand{\eq}{\eqref}

\def\bl#1{\renewcommand{\baselinestretch}{#1}} 
\def\cl{\centerline}
\def\bib#1{\bibitem[#1]{#1}}
\def\nt{\notag}
\def\bbx#1{ \boxed{\boxed{#1}}}

\def\lp{\left(}
\def\rp{\right)}
\def\lb{\left[}
\def\rb{\right]}
\def\la{\langle\, }

\def\ra{\,\rangle} 

\def\LB{\left\lbrace}
\def\RB{\right\rbrace}

\def\+#1#2{{#1+#2}}

\def\top#1#2{\smash{\mathop{\hbox to .2cm{$#2$}}\limits^{#1}}}

\def\0#1{{(#1)}}
\def\1#1{{\hat #1}}
\def\2#1{{\tilde #1}}
\def\3#1{{\boldsymbol#1}}
\def\4#1{{\,\mathbb#1}}
\def\5#1{{\mathcal#1}}
\def\6#1{_{\sst#1}}
\def\7#1{{\bar#1}}
\def\8{\infty}
\def\9#1{^{\,\sst#1}}
\def\/#1{{\bf#1}}
\def\;#1{{\breve#1}}
\def\`#1{{\mathring#1}}

\def\^{\wedge}

\def\bbh#1{{\mathbb{\widehat#1}}}

\def\bh#1{{\boldsymbol{\hat{#1}}}}
\def\bt#1{{\boldsymbol{\tilde#1}}}

\def\tc#1{\2{{\5#1}}}

\def\wh#1{{{\widehat#1}}}

\def\a{\alpha}

\def\c{\chi}
\def\d{\delta} 
\def\e{\varepsilon} 

\def\f{\phi} 
 
\def\g{\gamma}
\def\h{\eta} 

\def\k{\kappa}

\def\l{\lambda} 
\def\m{\mu} 
\def\n{\nu}
\def\o{\omega} 
\def\p{\pi} 

\def\q{\theta} 
 
\def\r{\rho}
 
\def\s{{\sigma}} 

\def\t{\tau} 
 
\def\x{\xi}
\def\y{\psi} 
\def\z{\zeta}

\def\D{\Delta} 
 
\def\G{\Gamma}
 
\def\O{\Omega} 

\def\Q{\Theta}

\def\Y{\Psi}

\def\hb#1{{\qq\text{#1}\qq}}
\def\bull{$\bullet\ $}
\def\db{{d\kern-.8ex {^-}}\!}

\def\Box{\square}
\def\={\equiv} 
\def\app{\approx} 
\def\cc#1{{{\mathbb C\hskip.5pt}^{#1}}}

\def\curl{\nabla\times} 
\def\div{\nabla\cdot } 
\def\eg{{\it e.g., }}

\def\grad{\nabla} 

\def\ie{{\it i.e., }}

\def\llra{\ \longleftrightarrow\ }
\def\iff{\ \Leftrightarrow\ }
\def\im{{\,\rm Im}\ }  
\def\imp{\ \Rightarrow\ }
\def\i1#1{\int_{-\infty}^\infty d#1\, } 
\def\inv{^{-1}}

\def\ir{\int_{-\infty}^\infty}

\def\lra{\leftrightarrow}

\def\mink{\4R^{3,1}}

\def\mt{{\,\mapsto\,}}

\def\pl{\partial}

\def\plra{\pl^{\kern-1.25ex^\lra}}
\def\qq{\quad} 
\def\qqq{\qquad} 
\def\qed{\vrule height6pt width4pt depth 0pt}
\def\re{{\,\rm Re}\  }   
\def\rr#1{{{\mathbb R}^{#1}}}

\def\sea{\searrow}
\def\sgn{{\,\rm Sgn \,}}
\def\sh#1{\hskip#1ex} 

\def\sr{\sqrt}

\def\supp{{\rm supp \,}}
\def\sv#1{\vskip#1ex}

\def\and{{\hbox{\  \ and\ \ }}}

\def\tf{\ \therefore\ }

\def\*{*\!}

\def\as{\int_{C_s} d\2k\ e^{ikz}}

\def\HB{\hfill\break}
\def\VE{\vfill\eject}

\begin{document}

\title{Physical wavelets and their sources:\\
Real physics in complex spacetime\thanks{\small Invited 
'Topical Review' for \sl  J.~Phys.~A: Math.~and  General, \tt    
www.iop.org/journals/jphysa \ \ 
\rm Preprint available at  \tt arxiv.org/abs/math-ph/0303027}}

\author{Gerald Kaiser\\
Center for Signals and Waves\\
www.wavelets.com}

\maketitle

\vspace*{-1.3 cm}

\bl{.6}
\tableofcontents

\VE

\bl1

\section{Overview}

 Since the introduction of electromagnetic wavelets in  \ci{K94}, I have been intrigued by two related questions:  (a) what are the sources responsible for their emission and absorption, and (b) can these sources be \sl realized \rm and used to launch and detect the wavelets? In principle, it should be easy to find the source for a known solution of a linear system. However,  these wavelets are \sl pulsed-beam propagators \rm that are, in a `poetic' sense at least,  waves generated by a single event  in \sl complex \rm spacetime as seen by an observer in \sl real \rm spacetime. Their sources, therefore, should be (ignoring polarization for simplicity)  \sl extended delta functions \rm $\2\d(x'-z)$, where $x'$ is the real observation point and $z$ the complex source point.  Attempts to make mathematical sense of all this have presented a number of challenges that have occupied me for the past several years. I am now pleased to report that the effort has been successful and the results are interesing. The sources for scalar (`acoustic') wavelets have been rigorously computed both in  spacetime and Fourier space and will be presented here.   The spacetime expressions  turn out to be singular objects with a wealth of geometric structure and, most interestingly,  their Fourier transforms are extremely simple --- contrary to all my previous expectations.  To understand how these simple expressions generate such rich spacetime structures, I complete the circle by computing the pulsed beams from their Fourier sources. As an unexpected byproduct, this gives an angular spectrum representation of  time-harmonic \sl complex-source beams, \rm used extensively in the engineering literature,  that  generalizes Hermann Weyl's well known representation of the fundamental solution for the Helmholtz equation.  The sources for \sl electromagnetic \rm wavelets are polarization and magnetization densities obtained by multiplying $\2\d(x'-z)$ by electric and magnetic dipole moments or, more generally, convolving them in time with variable dipoles representing a ` driving signal.'

  I believe that question (a) above has now been largely addressed, though some important points of interpretation  remain, as do no doubt many others.  This opens the possibility suggested by question (b).  Furthermore, although the wavelets were introduced in classical electrodynamics, I have hoped from the beginning that they may be useful in QED. This is encouraged by the simplicity of their momentum space propagators, which could offer a useful computational tool.

\section{Waves, wavelets, and complex spacetime}

I begin with a brief review of my past efforts to extend classical and quantum theories  to complex spacetime  and interpret the results physically.   By that I mean that the imaginary  spacetime coordinates, and any other extras associated with analyticity, are to be understood \sl directly \rm in terms of common observable attributes and not  merely as a technical device for proving theorems or exotic higher dimensions inaccessible to mortals stuck in the `real' world like the poor souls in Plato's cave. 

I have tried not to impose an \sl a priori \rm grand vision but, rather, 
interpret the imaginary coordinates in each theory by understanding their \sl effects \rm within that theory.  Consequently, the interpretations vary somewhat  from one theory to another.  But they all have in common the following theme. \sl  In the extended theory, certain singular points (evaluation maps on fields or wave functions, source points, etc.)  become `inflated' to extended objects. \rm     This transformation is determined  by analyticity and the particular  theory.  In every case, the structure of the objects is shaped  by the equations of the theory and their  degrees of freedom are specified precisely by the complex spacetime coordinates. The real coordinates give the \sl center,  \rm and the imaginary coordinates the \sl  extent  and orientation \rm  of the object in space and time.

These ideas are similar in  spirit  to wavelet analysis, where a function of one variable (time, say) is expressed in terms of an additional variable describing the scale or \sl resolution \rm in the first.  This analogy goes farther in the treatment of massless than massive fields, since the latter have an intrinsic scale and thus cannot be scaled arbitrarily. For relativistic fields with mass, spacetime `orientation' includes \sl velocity, \rm and this makes the complex spacetime an \sl extended phase space. \rm   The \sl relativistic coherent-state representations \rm for massive Klein-Gordon and Dirac fields constructed in \ci{K77, K78} (for single particles) and \ci{K87, K90} (for free quantized fields)  interpolate between `time-frequency' and `wavelet' descriptions, behaving like the former in the nonrelativistic regime and like the latter in the ultrarelativistic one. In fact, there is a very close correspondence between the nonrelativistic limit in physics and the \sl narrow-band approximation \rm in signal theory; see \ci{K90, K94, K96}.

Although  the results cited in this section are not new, I believe they have acquired some currency because of substantial progress recently in the understanding  of the  \sl sources \rm associated with \sl retarded \rm holomorphic\footnote{According to most referrences, \sl holomorphic \rm simply means \sl analytic. \rm  Possibly  the term is used to emphasize \sl complex \rm as opposed to real analyticity --- or perhaps just to intimidate the reader!
}
fields.  The new results focus on massless fields, but it is likely that similar computations exist for massive fields where the integrals are more difficult.  

Sources describe the breakdown of analyticity due to natural singularities and physically necessary branch cuts. What I find especially fascinating is that such branch cuts behave much like `real'  matter. Depending on the theory, they carry charge, mass and spin, and they emit and absorb radiation.  In spite of their simple origins, they turn out to have  surprising  and complex (pardon the expression) properties, the pursuit of which has the feeling of exploring  hitherto unknown forms of matter and not merely the mathematical properties of branch cuts.  The results of this search have intrigued and inpired me, and I hope to share this excitement with the reader. 

Partial  reports have appeared in \ci{K00, K01, K01a, K02, K02a, K03}, but the detailed computation of  complex spacetime point sources and their Fourier transforms, and the angular spectrum representation of complex-source beams,  has not appeared previously in print or preprint form.

\subsection{Spacetime and Fourier notation}

Real spacetime vectors will be written in the form of \sl complex Euclidean \rm vectors with real space coordinates and an imaginary time coordinate:
\begin{align}\lab{ict}
x=(\3x, it)\in M=\mink,\qq x^2=r^2-t^2,\qq r=|\3x|.
\end{align}
This is known in physics as  `$ict$' (we take $c=1$)  and is often regarded  to be unphysical  because it cannot be used  in  `generic' curved spacetimes where the metric tensor cannot be continued analytically in $t$. (See  \ci[p.~52]{MTW73}, \sl Farewell to ``$ict$.''\rm)   Therefore I feel compelled to explain at the outset why I use it nevertheless. The view of physical time as an essentially \sl imaginary \rm variable comes from the  idea of \sl complex distance \rm \ci{K00},  rooted in \sl Euclidean \rm spacetime,  which is the basis for  the recent progress in the analysis of sources.  In previous work  I had  followed a more conventional path, starting with known quantities in Minkowski space and extending them analytically when possible. That  approach, reviewed later in this section, works best for \sl free \rm fields, where the splitting into positive and negative frequency components provides a natural setting for analytic continuation. To understand sources, we must rather look at \sl propagators, \rm   \ie waves emitted by a single  \sl point  source \rm $\d\0x$, since waves emitted by all other sources can then be obtained by convolution.  However, propagators are much more complicated than  their elliptic counterparts,  the \sl potentials \rm due to point sources in Euclidean space. In the hyperbolic case, we must choose between retarded and advanced propagators (or hybrid ones, like Feynman's).  Each choice is singular not only at the source but also on at least one half of the light cone, and it is not clear whether and how such solutions can be extended  to complex spacetime.  The Euclidean case, by contrast, is a picture of simplicity. Because the $\d$-source in $\rr n$ is spherically symmetric, its potential  $G_n$ depends only on the distance $r$. To extend $G_n$ anlytically we need only extend  $r$, and this is easily done. The only complication is that the resulting \sl complex distance \rm $\2r\0z$ is double-valued in $\cc n$ because of  the square root in its definition, and a branch cut must be chosen.  For even $n\ge 4$, the extended  $G_n$ is even in $\2r$ and so does not depend on the choice of branch, while in all other cases it does.

This suggests an alternative strategy for extending physical fields without excluding the possibility of sources. Since the original signature is irrelevant once spacetime is made complex, why not \sl begin \rm in the  Euclidean setting, where operators are elliptic and life is simple, and continue analytically to the Minkowskian world? There may be  advantages to starting directly with Euclidean concepts, unencumbered by baggage imported from Minkowski space, even if in the end we intend to study propagators.   Thus we begin with the extended potential $G_4\0z$ in $\cc4$ and look for extended versions of the retarded  and advanced propagators $D\9+\0x$ and  $D\9-\0x$. However, since the concepts of propagation and causality are foreign to the Euclidean world,  $G_4\0z$ cannot distinguish between retarded and advanced propagators.  In fact, it turns out to be the extension of the \sl Riemann function, \rm which is the sourceless sum of the advanced and retarded propagators (point source). To split off  the retarded part, a branch must be chosen for the \sl spatial \rm complex distance $\2r_3$, representing the `distance'  from the complex source point at $\3z\in\cc3$ to the observer at $\3x'\in\rr3$. This gives a splitting into extended  propagators, $\2D\9+\0z$ and $\2D\9-\0z$  that depend on the branch cut  of $\2r_3$.  This branch cut,  $\5D\subset\rr3$, is  a `blown up' version of the original point source which  acts as a \sl disk source \rm for $\2D\9+\0z$ and 
$\2D\9-\0z$.\footnote{Going through the cut means entering a world where distance becomes negative and $\2D\9\pm\0z$ trade places.  As $\3z$ becomes real, the disk contracts to a point and the door is closed!}
Finally,   $D\9\pm\0x$ are recovered by an  `edge of the wedge' type  limit
\begin{align}\lab{edge}
D\9\pm\0x=\lim_{\e\sea0}\LB \2D\9\pm(x-i\e y)-\2D\9\pm(x+i\e y)\RB
\end{align}
with $y$ in the future cone of $\mink$.  This shows that while analytic extension is very simple for Euclidean potentials, it is  rather involved for propagators, lending support to the idea of  beginning in the Euclidean setting. But in this approach,  physical time must be imaginary,  just as `Euclidean time' must be imaginary when beginning with Minkowskian  or Lorentzian spacetime.  

At any rate, the dismissal of `$ict$' in \ci{MTW73} may have been premature. Even in general relativity, analytic continuations in time and space have borne some rich  fruit, even if the physical basis of the procedure is often ill-understood.   For example, an exquisitely simple geometric derivation of the Hawking temperature for Schwartzschild black holes is obtained \ci{HI79}  by analytically continuing the metric in time, interpreting the `Euclidean time' coordinate as an angle, and choosing its period to make the horizon a coordinate singularity like the origin in polar coordinates. The reciprocal of the imaginary time period  is interpreted in the usual (KMS) way as a temperature, and this turns out to be nothing but the Hawking temperature! I confess that I do not  understand this derivation in more than a formal way, but analytic continuation has, in any case, become the main strategy of black-hole thermodynamics, as explained in\ci{Kr03}.  The analytic continuation of spatial coordinates also has an honorable history in relativity, having played a major role  (and conceptually an equally obscure one) in the discovery of charged spinning black holes by Newman et al. \ci{N65}; see also \ci{N73, NW74, K01a, N02}.  And then there are the theories of twistors and H-spaces (see \ci{P87, HNPT78, BFP80} and references therein).

Returning to our summary of notations, we denote by  $M'$ the
  `momentum space' dual to $M$ and write the pairing between the two as the complex Euclidean scalar product,
\begin{gather}\lab{pair}
 k\cdot x=\3k\cdot\3x-\o t,\qq k=(\3k, i\o)\in  M',\  \  x=(\3x, it)\in M.
\end{gather}
The Fourier transform and its inverse are written
\begin{align}\lab{FT}
\1F\0k&=\int_M dx\  e^{-i k\cdot x} F\0x\\
F\0x&=\int_{M'} \db k\ e^{ik\cdot x} \1F\0k, \lab{IFT}
\end{align}
where we have introduced the notation 
\begin{align}\lab{vols}
 dx=d\3x\,dt \qqq \db k=\frac{d\3k\,d\o}{(2\p)^4} \qqq \boxed{\  \db=d/{2\p} \  }
 \end{align}
which eliminates most factors of $2\p$ in Fourier analysis. Strictly speaking, we should include factors of $i$ in front the 
integrals\footnote{To be precise, one should begin with dual vectors $(\3k, -i\o')$, since the frequency in units of $it$ is $-i\o'$,  so that $dx=id\3x dt$ and $\db k=-i\db\3k \db \o'$. The change to $\o=-\o'$ (or $k^4=-k_4$), which makes the phase velocity  
$+\3k/\o$, does not affect $\db k$ since $\ir \db \o'\cdots=\ir \db\o\cdots$.
}  
since the time and the frequency are imaginary, but this seems overly pedantic and we choose to leave them out, accepting the consequent possibility of  an extra factor of $i$ now and then.

\bf\sl Note: \rm Even though $x=(\3x, it)$ and $k=(\3k, i\o)$, we will write $F(\3x, t)$ and $\1F(\3k,\o)$ instead of $F(\3x, it)$ and $\1F(\3k, i\o)$ to keep the notation simple.

\sv1
We also need the following cones in $M$ and $M'$.

\bull The \sl positive and negative frequency light cones \rm 
\begin{gather}\lab{Cpm}
C\6\pm=\{(\3k, i\o)\in M':  \pm\o= \k >0\,\}
\hb{where} \boxed{\  \k \,=|\3k|\,. \ }
\end{gather}
Note that we exclude the `DC component' $k=0$, \ie nonvanishng constant solutions;  a wave must \sl oscillate \rm to be a wave.  (This is analogous to the admissibility condition in wavelet theory.) 

\bull The \sl future and past cones \rm 
\begin{align}\lab{Vpm}
V\6\pm=\{(\3y, iu)\in M: \pm u> a\} \hb{where} \boxed{\ a=|\3y| \ }
\end{align}
which are characterized by the duality relations
\begin{gather}\lab{dual}
 y\in V\6\pm\iff k\cdot y<0\qq \forall k\in C\6\pm\,.
\end{gather}
\bull The \sl solid positive and negative frequency cones \rm
\begin{align}\lab{Vpm2}
V'\6\pm\=\{k=(\3k, i\o)\in M': \pm\o\ge \k >0\}
\end{align}
which are the convex hulls of $C\6\pm$, characterized by
\begin{gather}\lab{dual2}
 k\in V'\6\pm\iff k\cdot y<0\qq \forall y\in V\6\pm\,.
\end{gather}

\bull The double cones
\begin{align}\lab{double}
C=C\6+\cup C\6-\,,\qq V=V\6+\cup V\6-\,,\qq V'=V'\6+\cup V'\6-\,.
\end{align}
Our main complex spacetime domains will the \sl forward and backward tubes \rm 
\begin{align}\lab{Tpm}
\5T\6\pm=\{z=x-iy: x\in M,\   y\in V\6\pm\} ,
\end{align}
and their union, which we call the \sl causal tube, \rm
\begin{align*}
\5T=\{z=x-iy: x\in M,\  y^2<0 \}.
\end{align*}

\subsection{The extension of relativistic fields}

All \sl free \rm relativistic fields extend analytically to $\5T$ in a sense to be explained. This can be done directly in spacetime by the  \sl analytic-signal transform \rm (AST)
 \ci{K90, KS92, K94}\rm\footnote{The AST introduced in the above references differs from \eq{ast} by a sign in order to be consistent with hyperfunction theory.  It will be seen to be Lorentz-invariant when applied to free fields.
} 
\begin{align}\lab{ast}
\2F(x-iy)=\frac1{2\p i}\ir \frac{ds}{s-iu}\ F(x-sy/u),\qq y=(\3y, iu)\in V.
\end{align} 
To see how this works,  suppose to begin with that $F$ is a general function not necessarily satisfying any differential equation, and  substitute  the Fourier expression for $F$:
\begin{align*}
\2F(x-iy)=\frac1{2\p i}\ir \frac{ds}{s-iu} \int_{M'} \db k\ 
e^{ik\cdot x}e^{-is k\cdot y/u}\ \1F\0k.
\end{align*}
Assuming the order of integration can be reversed, compute the integral over $s$  by closing the contour in the upper or lower half-plane depending on the behavior of the exponential. This gives
\begin{align}\lab{cauchy1}
\ir \frac{ds}{s-iu}\  e^{-is k\cdot y/u}= 
\1u\, \Q(-k\cdot y)\,e^{k\cdot y}
\end{align}
where $\Q$ is the Heaviside function and we have introduced a simple notation for the sign function, which will be used frequently: 
\begin{align}
\boxed{\  \1u=\sgn u\ } \hb{and} \Q\0\x=
\begin{cases}
1,& \x>0\\
0,& \x<0.
\end{cases}\lab{sgn}
\end{align}
(This is easily remembered since $\1u$ just a 1D version of  a unit vector $\bh u$.)  That gives the AST as an extension of the inverse Fourier transform:
\begin{align}\lab{ast2}
\2F(x-iy)=\1u \int_{M'} \db k\  \Q(-k\cdot y)\,e^{ik\cdot (x-iy)} \,\1F\0k,
\end{align} 
which also tells us that the Fourier transform in $x$ of the AST is
\begin{align}\lab{hatast}
\1F(k,y)\=\int_M dx\ e^{-ik\cdot x}\, \1F(x-iy)
=\1u\, \Q(-k\cdot y)\,e^{k\cdot y} \,\1F\0k.
\end{align}

In general $\2F$ is  not  analytic, in spite of the notation, because the Heaviside factor spoils analyticity as $y$ varies. ($F$ may not have an analytic extension.) However, if $\1F\0k$ is supported in $V'$ \eq{Vpm2}, then the Heaviside function in \eq{ast2} disappears after correlating positive frequencies with the future cone and negative frequencies with the past cone:
\begin{align*}
\Q(-k\cdot y)=\Q(\o u)=
\begin{cases}
1,&  k\in V'\6\pm \hb{and} y\in V\6\pm\\
0,&  k\in V'\6\pm \hb{and}  y\in V\6\mp
\end{cases}
\end{align*}
and we get a third expression for the AST,
\begin{align}\lab{ast3}
\2F\0z=\pm\int_{V'\6\pm}\db k\ e^{ik\cdot z}\, \1F\0k ,\qqq z\in\5T\6\pm\,.
\end{align}
Now
\begin{align*}
k\in V'\6\pm\,,\  \   y\in V\6\pm\imp k\cdot y\le \k (a\mp u)<0,
\end{align*}
hence the factor $e^{k\cdot y}$ in the Fourier-Laplace kernel 
$e^{ik\cdot z}$ decays exponentially. If $\1F\0k$ does not \sl grow \rm exponentially, the integrals \eq{ast3} define holomorphic functions in $\5T\6\pm$. (Note, however, that the exponential decay gets weaker and weaker as $y$ approaches the light cone. This will later give the ability to \sl focus \rm pulsed-beam wavelets.)

Of course, it cannot be claimed that $\2F$ is an extension of $F$ itself since only the positive-frequency part is represented in $\5T\6+$ and
only the negative-frequency part in $\5T\6-$.   Define the \sl partial \rm boundary values
\begin{align*}
\2F(x\mp i0)=\lim_{\e\sea 0} \2F(x\mp i\e y),\qqq y\in V\6+
\end{align*}
which do not depend on the particular choice of $y$. Then  $F$ is a  boundary value of $\2F$ in the same sense as  \eq{edge},\footnote{The \sl sum \rm gives a spacetime version of the Hilbert transform \ci{K90}.}
\begin{align}\lab{edge2}
F\0x=\2F(x-i0)-\2F(x+i0).
\end{align}
Generally, the restrictions of $\2F\0z$ to the disjoint domains $\5T\6\pm$ are unrelated holomorphic functions, but if $F\0x$ vanishes in an open region of spacetime, then \eq{edge2} implies that they are part of a \sl single \rm holomorphic function. This is the famous \sl edge of the wedge \rm theorem \ci{SW64}.

It is instructive to compute the AST of a function depending only on \sl time: \rm 
\begin{align}
\2F(t-iu)&=\frac1{2\p i}\ir  \frac{ds}{s-iu}\, F(t-s),\qqq u\ne 0\notag\\
&=\frac1{2\p i}\ir  \frac{ds}{t-iu-s}\, F(s)\notag\\
&=\1u\ir \db\o\  \Q(\o u)\,e^{-i\o(t-iu)} \,\1F\0\o\lab{ast4}
\end{align}
or
\begin{align}\lab{AS}
\2F(t-iu)=\begin{cases}
\qq  \int_0^\8\db\o\ e^{-i\o(t-iu)} \,\1F\0\o, &u>0\\
 - \int_{-\8}^0\db\o\ e^{-i\o(t-iu)} \,\1F\0\o, &u<0.
\end{cases}
\end{align}
This is a pair of  \sl analytic signals \rm  extending the positive and negative frequency parts to  the lower and upper complex half-planes,
a useful concept  introduced by Dennis Gabor \ci{G46} which also explains the name of our transform.  (Actually, Gabor worked with real signals, where it suffices to consider only one of the above pair since the other is merely its complex conjugate.)

Equation \eq{ast4}  shows that $\2F$ is a convolution of $F$ with the \sl Cauchy kernel, \rm
\begin{align}\lab{conv3 }
\2F\0\t=\ir ds\ C(\t-s) F\0s,\qq \t=t-iu,\qq C\0\t=\frac1{2\p i\t}\,,
\end{align}
and that the  Fourier transform of $C$ is
\begin{align}\lab{hatC}
\1C(\o, u)\=\ir dt\ e^{i\o t} C(t-iu)=\1u\,\Q(\o u) e^{-\o u}.
\end{align}
The role of the sign $\1u$ in the AST can be understood by assuming
that $F\0t$ is compactly supported in an interval $I$.  Then, according to the edge of the wedge theorem, $\2F\0\t$ is holomorphic for all $\t\notin I $ and  we may therefore rewrite \eq{ast4}  as a contour integral
\begin{align}\lab{ast5}
\2F\0\t&=\frac1{2\p i}\int_\g  \frac{ds}{\t-s}\, \2F\0s,\qq \t=t-iu,
\end{align}
where $\g$ is a closed contour surrounding $I$,  running in the \sl positive \rm direction at $s-i0$ and in the \sl negative \rm direction at $s+i0$.  Equation \eq{ast5} is Cauchy's formula for the values of an analytic function \sl outside \rm of $\g$ if the contour can be deformed through infinity in the Riemann sphere.  This interpretation would not exist without the sign in the definition of the AST. 

Equation \eq{ast5} shows that the sign $\1u$ in \eq{ast4}  gives a  positive orientation  to the boundary of the lower-half complex time plane (\ie the time axis as seen from below) and  a negative orientation  to the  boundary of the upper-half complex time plane (the time axis as seen from above), which is obviously correct.  This interpretation carries over directly to the spacetime AST.  In a reference frame where $y=(\30, u)$, \eq{ast} reduces essentially to \eq{ast4} :
\begin{align*}
\2F(\3x, \t)=\frac1{2\p i}\ir  \frac{ds}{\t-s}\, F(\3x, s),\qq \t=t-iu.
\end{align*}

In the sequel, it will be helpful to keep in mind the following correspondence between the 4D geometry of spacetime and the 1D geometry of time:
\begin{align*}
V'\6\pm\,,\ C\6\pm &\llra   \{\o: \pm \o>0\}\\
V\6\pm &\llra \{u: \pm u>0\}\\
\5T\6\pm & \llra \{ t\mp iu: \pm u>0\}.
\end{align*}

\subsection{Massive fields and relativistic coherent states}

Free particles and fields of mass $m>0$ satisfy the Klein-Gordon equation
\begin{align}\lab{KG}
\Box F\0x\=(\D-\pl_t^2)F\0x=m^2 F\0x.
\end{align}
(In the case of spinor fields, all components satisfy \eq{KG}. We consider scalars for simplicity. Dirac particles and fields are treated in \ci{K87, K90}.)  In Fourier space, this means
\begin{align*}
& (k^2+m^2)\1F\0k=0\\
\imp \1F\0k&=2\p\d(m^2+k^2)\,f\0k=2\p \d(E^2-\o^2)\, f\0k,\qq E=\sr{m^2+\k ^2}\\
&=\frac\p E\LB \d(E-\o) f(\3k, E)+\d(E+\o) f(\3k,-E)\RB
\end{align*}
for some function $f$ defined on the \sl double mass shell \rm
\begin{align*}
\O\6\pm=\{k\in M': \pm\o=E\},\qq \O=\O\6+\cup\O\6-\,.
\end{align*}
Therefore
\begin{align*}
F\0x=\int_{\O\6+} d\2k\ e^{ik\cdot x} f\0k
+\int_{\O\6-} d\2k\ e^{ik\cdot x} f\0k=\int_\O d\2k\ e^{ik\cdot x} f\0k,
\end{align*}
where
\begin{align}\lab{dkm}
d\2k=\frac{\db\3k}{2E}=\frac{d\3k}{16\p^3 E}
\end{align}
is the Lorentz-invariant relativistic measure on $\O\6\pm$ \ci{IZ80}. Applying the AST gives
\begin{align*}
\2F\0z&=\1u\int_\O d\2k\ \Q(-k\cdot y)\,e^{ik\cdot z} f\0k\\
&=\pm\int_{\O\6\pm} d\2k\ e^{ik\cdot z} f\0k,\qqq z=x-iy\in\5T\6\pm\,.
\end{align*}
This has been used to build a coherent-state representation for massive particles (where $f\0k$ is a `c-number'  function vanishing on $\O\6-$ to give a positive-energy solution) and  fields (where $f(\3k, E)$ and $f(\3k, -E)$ are creation and annihilation operators).  These representations have a physical interpretation relating $y$ to the \sl  expected  energy-momentum \rm of the coherent states. I am not going to repeat this construction here as it is readily available \ci{K77, K78, K87, K90}, but merely indicate how the above interpretation comes about.

Fix $y\in V\6+$ and consider the exponential $e^{k\cdot y}$ as a function of $k\in\O\6+$.  The Lorentzian  scalar product satisfies the reverse Schwartz inequality
\begin{align*}
k\cdot y\le-m\l,\qq \hb{where} \l=\sr{-y^2}=\sr{u^2-a^2}>0
\end{align*}
which becomes an equality if and only if $k$ is parallel to $y$:
\begin{gather}\lab{ky}
k\cdot y= -m\l  \iff k=(m/\l) y\=k_y\\
\tf e^{k\cdot y}\le e^{-m\l} \hb{and}  
e^{k\cdot y} = e^{-m\l} \iff k=k_y\,.\notag
\end{gather}
Therefore $e^{k\cdot y}$ acts as a \bf\sl ray filter \rm in momentum space, favoring those plane waves propagating approximately in the direction of $y$. The larger we take $\l$, the stronger the filter and the more collimated the ray bundle passed by it.  The coherent states are defined in momentum space by
\begin{align}\lab{ez}
e_z\0k=e^{-ik\cdot z^*}, 
\end{align}
so that they act as \sl evaluation maps \rm on the Hilbert space $\tc H$ of holomorphic solutions with inner product defined in $L^2(d\2k)$:
\begin{align}\lab{eval}
\2F\0z=\la e_z| f\ra,\qqq \la \2F_1 | \2F_2\ra\=\la f_1 | f_2\ra.
\end{align}
This makes $\tc H$ a  reproducing kernel Hilbert space, and that 
kernel is
\begin{align}\lab{rk }
K(z', z^*)\=\la e_{z'}\,|\,e_z\ra=\int_{\O\6+}d\2k\ e^{ik\cdot(z'-z^*)}
=-i\D\9+(m;z'-z^*),
\end{align}
where $\D\9+$ is the \sl Wightman 2-point function for the scalar field of mass $m$. \rm

To obtain a resolution of unity, choose any oriented 3-dimensional submanifold  $S$ in $M$ as a prospective configuration space and the 6-dimensional submanifold
\begin{align*}
\s_\l=\{ x-iy\in\5T\6+: x\in S,\  y^2=-\l^2\}, \qqq \l>0
\end{align*}
as the associated  classical phase space, with $\l>0$ arbitrary.  A symplectic form and  covariant measure on $\s_\l$ are chosen as follows. Define the invariant 2-form and 6-form
\begin{align*}
\a=dx_\m\^ dy^\m,\qqq \a^3=\a\^\a\^\a.
\end{align*}
Then the following are proved:

\bull The restriction of $\a$ to $\s_\l$ is a symplectic form if and only if $S$ is \sl nowhere timelike, \rm \ie it must be locally  spacelike or lightlike. In other words, \sl $\s_\l$ is a reasonable phase space if and only if $S$ is a reasonable configuration space, \rm so the symplectic geometry is compatible wth the spacetime geometry.

\bull If $S$ satisfies the above condition, then $\a^3$ defines a positive measure $d\m_\l$ on $\s_\l$ and we have  a resolution of unity
\begin{align}\lab{ru}
\int_{\s_\l} d\m_\l\0z\ |e_z\ra \la e_z|=I,\qqq d\m_\l=C(\l, m)\, \a^3
\end{align}
where $C(\l, m)$ is an invariant and $I$ is the identity operator on $L^2(d\2k)$.

\bull The physical interpretation of $\s_\l$ as a phase space is confirmed explicitly in the case when $S$ is flat by the expected positions and momenta in the state $e_z$,
\begin{align}\lab{Pm}
&\la P_\m\ra_{e_z}=a\,y_\m && y_\m=-\im z_\m && \m=0, 1,2,3\\
&\la  X_j(t)\ra_{e_z}=x_j &&  x=(\3x, it)  && j=1,2,3\notag
\end{align}
where $a(\l, m)$ is an invariant and $X_j$ are the Newton-Wigner operators in the Heisenberg picture at time $t$. (When $S$ is \sl curved, \rm the  positions operators obtained by quantization on $\s_\l$ do not commute; see  \ci{K76}.)

\bull The parameter $1/\l$ measures the uncertainty or \sl resolution \rm  in the momentum of coherent states parameterized by $z\in\s_\l$, in accordance with the above discussion of ray filters.

\bull  Unlike the usual spacetime representation,  \sl the coherent-state  representation admits a conserved, covariant probability current density, \rm given by
\begin{align}\lab{current}
j_\m\0z=-\frac{\pl  |\2F\0z|^2}{\pl y^\m}
=i\2F\0z^*\,\frac{\pl  \2F\0z}{\pl x^\m}
-i\2F\0z\,\frac{\pl  \2F\0z^*}{\pl x^\m}.
\end{align}

\bull In the nonrelativistic limit one  obtains a coherent-state representation of the centrally extended Galilean group with Gaussian measure in momentum:
\begin{align}\lab{NR}
d\m_\l\9{\rm NR}= C\,e^{-m\3y^2/u}\ d\3x\,d\3y,\qqq  
\3y=\frac u m \la \3P\ra.
\end{align}
The weight function is imposed on the relativistic  (hence  \sl uniform\rm)  measure $d\m_\l$ in  compensation for the  deformation of the mass shell  $\O\6+$ to a 3-plane at infinity \rm (the nonrelativistic momentum space). Upon applying a `holomorphic gauge transformation' \ci{KM80} to the nonrelativistic wave functions (solutions of Schr\"odinger's equation), the  representation becomes identical to  the  Bargmann-Segal representation  of the Weyl-Heisenberg group:
\begin{align} \lab{NR2}
\2F\6{\rm NR} \to e^{m\3z^2/4 u}\2F\6{\rm NR}\imp 
d\m_\l\9{\rm NR} \to C\, e^{-m|\3z|^2/2u}\, d\3x\, d\3y.
\end{align}

\subsection{Massless fields and wavelets}

In the limit $m\to 0$, the 'reconstruction constant' $C(\l, m)$ in \eq{ru} diverges and the coherent-state representation is no longer 
square-integrable due to disappearance of  the `mass gap'   
$\k\ge m>0$. At the same time,  the symmetry group grows from the Poincar\'e group to the conformal group $\5C$.  In its realization as $SU(2,2)$, $\5C$ acts on $\5T\6\pm$ by matrix-valued M\"obius transformations. Thus it is reasonable to look for resolutions of unity  adapted to the new symmetries. This took me several years to realize, and only when studying wavelet theory in preparation for Ingrid Daubchies'  1990 \sl Ten Lectures \rm conference \ci{D92} did I understand that \sl scaling \rm needed to be brought into the picture. Since massless fields are important in classical as well as quantum physics, I decided to begin with classical fields, the prime examples of which are acoustic and electromagnetic fields. This led to the construction of acoustic and electromagnetic wavelets \ci{K92, K94, K94a}.   From a foundational as well as applied point of view, I believe the electromagnetic wavelets hold far more promise and the `acoustic' ones serve mainly to simplify the analysis by stripping away all complications related to polarization.  I now briefly review the construction of acoustic wavelets, leaving the electromagnetic ones to the end of the paper where they and  their sources  will be constructed from the  scalar acoustic sources.

Solutions of the scalar wave equation are given by
\begin{align}\lab{acous}
\Box F\0x =0\imp F\0x=\int_C d\2k\ e^{ik\cdot x} f\0k,
\end{align}
where $d\2k=\db\3k/2\k $ is the massless version of \eq{dkm} on the light cone $C$ \eq{double}. Applying the AST gives the extension to $\5T\6\pm$,
\begin{align}\lab{acous2}
\2F\0z&=\1u\int_C d\2k\ \Q(-k\cdot y) e^{ik\cdot z}\,f\0k
=\pm\int_{C_\pm}d\2k\  e^{ik\cdot z}\,f\0k.
\end{align}
In the massive case, we chose the coherent states to be the complex conjugates of $ e^{ik\cdot z}$. However, this cannot be done now because $e^{ik\cdot z}$ does not vanish near $\o=0$ and this will spoil the resolution of unity.  (In wavelet terms,  
$e^{ik\cdot z}$ is not \sl admissible.\rm)
We get around this difficulty by changing the inner product  of solutions to
\begin{align}\lab{IPnu}
\la F_1| F_2\ra\=\la f_1| f_2\ra
=\int_C\frac{d\2k}{\k ^\n}\ f_1\0k^* f_2\0k, \qq \n\ge 0.
\end{align}
For $\n=0$ this is the Lorentz-invariant inner product, but it will turn out that we need $\n>1$ to obtain admissible wavelets.  Next, write the extension \eq{acous2} in the form of an inner product
\begin{align}\lab{acous3}
\2F\0z=
\1u\int_C \frac{d\2k}{\k^\n}\  \k ^\n\, \Q(-k\cdot y) e^{ik\cdot z}\,f\0k=
\la \y_z| f\ra,
\end{align}
where
\begin{align}\lab{yz }
\y_z\0k=\1u\,\k ^\n\, \Q(-k\cdot y)\, e^{-ik\cdot z^*}
\end{align}
are to be the `acoustic wavelets' in Fourier space, representing spacetime solutions
\begin{align}\lab{Yx}
\Y_z(x')=\1u\int_Cd\2k\ \k ^\n\, \Q(-k\cdot y)\, e^{ik\cdot (x'- x+iy)}
\=\Y(x'-z^*).
\end{align}
From the invariance of $d\2k$ under  $k\to -k$ it follows that 
\begin{align}\lab{Yx2}
\Y(-z)=-\Y\0z,
\end{align}
hence it suffices to compute $\Y$ in $\5T\6+$. 
For any fixed $y\in V\6+$, the function $\Y(x-iy)$ may be called a `mother wavelet'  from which all others are obtained by complex translations. Notice that \sl imaginary translations \rm  generalize scaling. \rm

As in the massive case, where  a great deal of freedom existed to choose  a phase space $\s$ due to the abundance of coherent states,  there is now a lot of freedom in choosing a family of $\Y_z$s to build a resolution of unity. Perhaps the simplest choice for a \sl continuous \rm frame is by analogy with 1D wavelets, which are parameterized by position and scale. `Position' is now $\3x$, and we take `scale' to be $u$, since it dominates the other scale parameters $\3y$. Thus we fix any time $t$, say $t=0$, and set $\3y=\30$ (this will give \sl spherical \rm wavelets). Our parameter space is then
\begin{align}\lab{EST}
E=\{z=(\3x, -u): \3x\in\rr3,\ \ u\ne 0\},
\end{align} 
which is \sl Euclidean spacetime \rm consisting of real space and imaginary time coordinates, with the Euclidean time $u$ acting as a scale  in $\mink$.  A quick  dimensional analysis shows that to
 compensate for the weight $\k^\n$ in Fourier space, we need the measure
\begin{align*}
d\m_\n\0z= C_\n \,d\3x\,|u|^{\n-2} du,
\end{align*}
with $C_\n$ adjusted to give a resolution of the identity $I_\n$ in $L^2(d\2k/\k^\n)$:
\begin{align}\lab{ru2}
\int_E d\m_\n\0z \,|\Y_z\ra \la \Y_z|=I_\n,\qqq C_\n =2^\n/\G(\n-1).
\end{align}
Thus we must take  $\n>1$ to get  `admissible' wavelet representations,
and  \eq{ru2} then represents solutions of the wave equation as superpositions of spherical wavelets centered at $\3x$ with a pulse duration  $T=\5O\0u$. These wavelets are sourceless, \sl converging \rm onto $\3x$ when $t<0$ and \sl diverging \rm from $\3x$ when $t>0$, and their radius at the \sl waist \rm $t=0$ is, like the pulse duration,  $R=\5O\0u$. As promised, $u$ controls all scales.

Converging  spherical wavelets are unnatural under ordinary conditions.\footnote{There are exceptions, including the recent method of \sl time-reversed acoustics \rm where sound is recorded by an array of microphones which are then  played in reverse, sending the waves back.
}
We want  to eliminate the converging (advanced) part and retain only the diverging (retarded) part.  The resulting wavelets will have \sl sources, \rm  but the splitting cannot be done by brute force (\eg multiplying by the Heaviside function) since that will spoil the analyticity and amount to introducing sources with infinite support. A natural separation into advanced and retarded wavelets was found in \ci[Chapter 11]{K94} while computing $\Y\0z$. We review this because it foreshadows the recent developments.  

Assuming $\n$ is a nonnegative integer, we have for $z\in\5T\6+$
\begin{align}\lab{Gz}
\Y\0z&=\int_{C\6+} d\2k\ \k^\n \,e^{ik\cdot z}=(-\pl_u)^\n G\0z\\
G\0z&=\int_{C_+} d\2k\ e^{ik\cdot z}=-\frac{1}{4\p^2z^2},\notag
\end{align}
where $G\0z$, the original inadmissible kernel with $\n=0$, is most easily computed by using the Lorentz-invariance of the integral. Now
\begin{align*}
 -z^2=(t-iu)^2-(\3x-i\3y)^2=(t-iu-\2r)(t-iu+\2r)
\end{align*}
where
\begin{align}\lab{cd}
\2r(\3x-i\3y)=\sr{(\3x-i\3y)^2}, \qqq \re\2r\ge 0
\end{align}
is the \sl complex distance \rm from the source point $i\3y$ to the observation point $\3x$, which will be studied in detail below. Thus $G\0z$ can be expanded  in partial fractions, giving
\begin{align}
G\0z&=\frac{1}{8\p^2\2r} \LB \frac1{t-iu-\2r}-\frac1{t-iu+\2r}\RB,\qqq z\in\5T\6+\notag\\
\Y\0z&=\Y\9+\0z+\Y\9-\0z\notag\\
\Y\9\pm\0z
&=\pm\frac{i\G(\n+1)}{8\p^2\2r}\, \frac1{(u+ i(t\mp\2r))^{\n+1}}\,.
 \lab{split}
\end{align}
The expression for $\Y\0z$ remains valid if $\n$ is  not  an integer, provided an appropriate branch cut is chosen. It splits naturally into retarded and advanced parts without spoiling holomorphy \sl everywhere, \rm as would a brute-force splitting. Instead, the two parts acquire singularities from the complex distance function. For given $\3y\ne \30$,

\bull  they diverge on the \sl branch circle \rm $\5S$ of radius $|\3y|$ in the plane $\3y\9\perp$,  where  $\2r=0$;

\bull they are discontinuous across the \sl branch disk \rm $\5D$ spanning $\5S$;

\bull  $\5D$ is a \sl source \rm for $\Y\9\pm$, canceling in $\Y$.

\bull  $\Y\9-$ and  $\Y\9+$ are \sl pulsed beams \rm converging to, then diverging from $\5D$.

We have called $\2r$  the `distance' from an imaginary source point $i\3y$ to a real observation point $\3x$. At this stage, such language must be viewed as `poetry' since the idea of a point source at $i\3y$ has not been defined and it is not even clear what it means.  In physics, a complex distance function identical to $\2r$ was the basis for the construction of  spinning, charged black holes; see  \ci{N65, N73}. In engineering,  `complex-source pulsed beams' similar 
to the above\footnote{I was unaware of this work when developing acoustic and electromagnetic wavelets, and thank Lou Felsen for pointing it out.
}
have been applied extensively since the 1980s, and their time-harmonic components, known as 'complex-source beams,'  since 
the 1970s; see \ci{HF01} for a comprehensive review.  But until recently,  no serious study seems to have been undertaken to  make mathematical sense of the idea of a ``complex point source,'' whether time-harmonic or pulsed, and therefore of how such beams may be realized.  (See \ci{HLK00} for an early attempt.) Perhaps this is because their singular and convoluted structure in the near zone appears to make the requisite analysis difficult if not impossible.  We will see that the sources are tractable in spacetime and, most significantly, even simple and computationally effective in Fourier space.

Although the requirement $\n>1$ precludes a Lorentz-invariant wavelet representation for acoustic waves, it does admit one for electromagnetic waves, where the invariant measure on $C$ has $\n=2$. Note that \begin{align}\lab{dmu3}
\frac{d\2k}{\k^2}=\frac{\db\3k}{2\k^3}
\end{align}
is \sl scaling invariant \rm as well as Lorentz-invariant.  In fact, the associated Hilbert space of solutions of Maxwell's equations carries a unitary representation of the full conformal group $\5C$, as proved by
Gross \ci{Gr64}. The electromagnetic  wavelet representation is likewise  covariant under $\5C$ \ci[Chapter 9]{K94}, and this opens up some interesting applications.  I will not discuss the details here since  more recent developments are discussed later.

\section{Point sources in complex space}

We now begin implementing the `Euclidean strategy' of setting up base camp in the Euclidean world from which to tackle the hyperbolic world by

\bull analytically extending the fundamental solutions of Laplace's  equation, 

\bull computing the extended  $\d$-sources, and  

\bull extracting Minkowskian propagators and sources from these extensions.

Because we will need the  fundamental solutions in $\rr3$ as well as $\rr4$, we work in this section with $\rr n$ for $n\ge 3$. The fundamental solution \ci{T96} $G_n$ for Laplace's equation in $\rr n$ is
\begin{gather}\lab{Lap}
\D_n G_n(\3x)=-\d_n(\3x), \qq  G_n(\3x)=\frac1{\o_n}\, \frac{r^{2-n}}{n-2}, \qq n \ge 3,
\end{gather}
where $\o_n=2\p^{n/2}/\G(n/2)$ is  the area of the unit sphere in $\rr n$ and
\begin{align*}
r(\3x)=\sr{\3x^2}
\end{align*}
is  the Euclidean distance.\footnote{The potential in $\rr2$ is 
$G_2=-(2\p)\inv\ln r$. Although our method works  here as well, this case is somewhat special and will be treated elsewhere \ci{K0x}.}
To extend $G_n$ analytically, we need only extend $r$. Define the 
\sl complex distance \rm from  $i\3y$ (source point) to $\3x$ (observation point)  as
\begin{align}\lab{cd2}
\2r(\3x-i\3y)=\sr{(\3x-i\3y)^2}=\sr{r^2-a^2-2i\3x\cdot\3y}.
\end{align}
Fixing $\3y\ne\30$, the branch points form an 
$(n-2)$-sphere in the hyperplane  $\3y\9\perp$ of $\rr n$,
\begin{align}\lab{Sy}
\5S(\3y)=\{\3x\in\rr n:  r=a, \   \3x\cdot\3y=0\}.
\end{align}
 We will use the cylindrical coordinates\footnote{To help visualize the geometry, the reader may think of the case $n=3$, where $\5S(\3y)$ is the circle forming the equator  of the sphere whose  north pole  is $\3y$ and the unit vector $\3\s$ is  represented by the azimuthal angle  $\f$ in the  plane $\3y\9\perp$.}  $(\r, \3\s, \x)$  given by
\begin{align}\lab{cyl}
\x=\bh y\cdot \3x,\qq \r=\sr{r^2-\x^2}\qqq (\bh y=\3y/a), 
\end{align}
with $\3\s$ a vector in the \sl unit \rm sphere $\5S(\bh y)$ of the hyperplane $\3y\9\perp$. The intersection of $\5S(\3y)$ with the half-plane through $\3y$ and $\3\s$ is the point $(a, \3\s, 0)$. If $\3x$ follows a simple loop surrounding  this point (\ie if the observer  \sl links \rm the circle), then $\2r$ changes sign. To make $\2r$ single-valued, we must prevent the completion of such loops by choosing a \sl branch cut \rm consisting of a hypersurface  with $\5S$ as its boundary. The branch cut must be chosen so that $\2r$ reduces to the usual distance on $\rr n$,
\begin{align*}
\3y\to\30\imp\2r(\3x-i\3y)\to +r(\3x).
\end{align*}
The simplest such cut  is obtained by requiring
\begin{align*}
\re \2r\ge 0,
\end{align*}
which gives  the \sl disk \rm spanning $\5S(\3y)$,
\begin{align*}
\5D(\3y)=\{\3x:  r\le a, \   \3x\cdot\3y=0\},\qq \pl\5D=\5S.
\end{align*}
The most general branch cut is a \sl membrane \rm obtained from $\5D$ by a continuous deformation leaving its boundary intact.

\bf\sl Note: \rm By `branch cut' we really mean a \sl slice \rm of the branch cut of $\2r(\3z)$ at constant $\3y$, since the source is taken as fixed.

Fixing $\3y\ne\30$, write
\begin{align}\lab{pq}
\2r\=\sr{(\3x-i\3y)^2}=p(\3x) -iq(\3x)
\end{align}
with the $\3y$-dependence understood. Then $p(\3x)\ge 0$ in $\rr n$,   $\5D$ is characterized by $p=0$, and $\5S$ by $p=q=0$.
Squaring \eq{pq}  gives
\begin{align}
&r^2-a^2-2i \3x\cdot\3y=p^2-q^2-2ipq\notag\\
\imp& r^2-a^2=p^2-q^2,\qq  \3x\cdot\3y=pq.  \lab{r2}
\end{align}
The cylindrical coordinates \eq{cyl} are thus given by
\begin{align}
a \x&=pq\notag\\ 
a^2\r^2&=a^2r^2-p^2q^2\notag\\
&=a^2(p^2-q^2+a^2)-p^2q^2 \notag\\  
&=(p^2+a^2)(a^2-q^2). \lab{cyl2}
\end{align}
In particular, note that $|q|\le a=|\3y|$, \ie \sl the imaginary part of $\2r(\3z)$ is bounded by the modulus of the imaginary part of $\3z$. \rm

\sv4
\cl{\includegraphics[width=6.5 in]{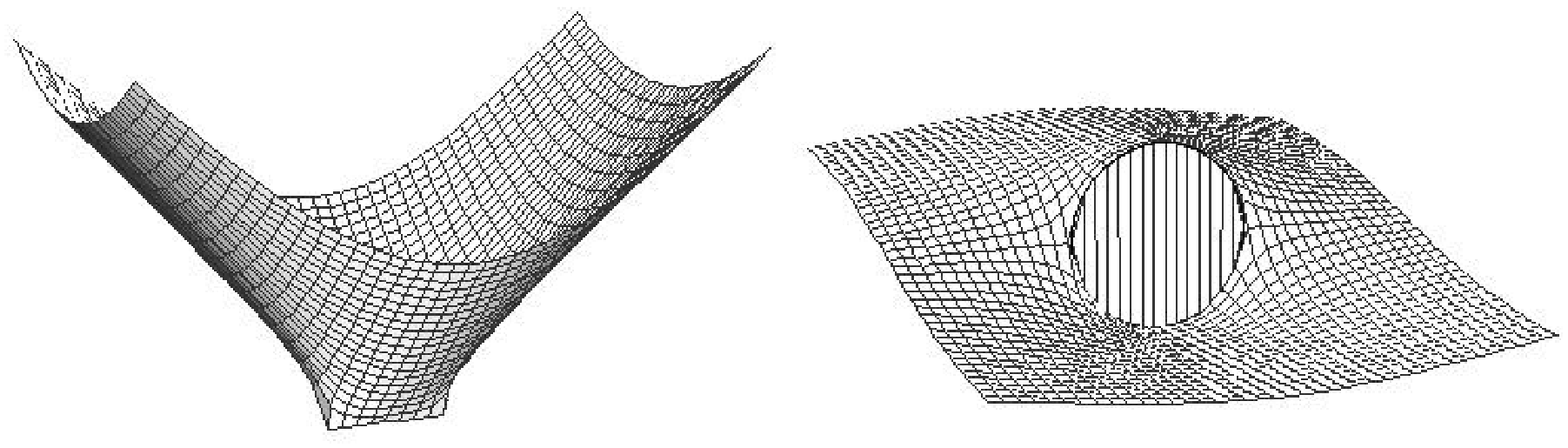}}
\nopagebreak[4]
{\small \bf Figure 1. \rm Plots of Re $\2r$ (\sl left\rm) and Im $\2r$
 (\sl right\rm) with $n=2$ and $\3y=(0, 1)$.   $D(\3y)$  is now the interval  $[-1, 1]$ along the $x_1$ axis. The graph of $\re \2r(\3x)$ is a pinched cone, and the jump of $\im\2r(\3x)$ across $D(\3y)$ is $2\sr{1-x_1^2}$, hence the cut is circular.  Note from \eq{far} that far from the disk, $\re\2r\app r$ and $\im\2r\app -\cos\q$ (since $a=1$ here).}
\sv2

It follows immediately from \eq{cyl2} that the \sl level surfaces \rm of $p(\3x)$ and $q(\3x)$ are 
\begin{gather}
E_p\=\{\hbox{constant } p>0\}=\LB\3x: \frac{\r^2}{p^2+a^2}+\frac{\x^2}{p^2}=1\RB\lab{level}\\
H\9\pm_q\=\{\hbox{constant } 0<\pm q<a\}
=\LB\3x: \frac{\r^2}{a^2-q^2}-\frac{\x^2}{q^2}=1, \  \  \pm \x>0\RB.\notag
\end{gather}
The $E_p$'s are a family of \sl oblate spheroids \rm filling the complement of $\5D$ in $\rr n$,  with $\5S$ as their common focal set, and the $H\9\pm_q$'s are the orthogonal family of upper and lower semi-hyperboloids, also $\5S$-confocal and  joining in $\5D$.  As $p\to 0$, $E_p$ converges to a \sl double cover \rm of $\5D$, a fact that will be important in our computations. Similarly, as $q\to 0$, $H\9\pm_q$ converge to the upper and lower  covers of the \sl complement \rm of $\5D$ in the hyperplane $\x=0$. Finally, as $q\to\pm a$, the semi-hyperboloids collapse to half-lines:
\begin{align*}
H_a\9\pm=\{q=\pm a\}=\{\pm \l\3y: \l>0\}.
\end{align*}
For the pulsed beams, the $E_p$'s (with $n=3$) will be \bf\sl wave fronts, \rm the  $H\9\pm_q$'s give the orthogonal surfaces of  \bf\sl radiation flow, \rm and $H_a\9\pm$ will be the forward and backward \sl beam axes. \rm  Use will also be made of the  \sl far zone approximation, \rm where the observer is far from the disk. By \eq{r2},
\begin{align}\lab{far}
r\gg a\imp p\app r,\  \  q\app a\cos\q
\end{align}
so that $E_p$ becomes the  sphere $r=p$ and $H\9\pm_q$ become the  cones  $\cos\q=\pm q/a$.

The complex distance thus provides a natural set of coordinates in $\rr n$, called \sl oblate spheroidal \rm (OS) \sl coordinates,  \rm  given by
\begin{align*}
(p, q, \3\s):\ p\ge 0,\  -a\le q\le a,\  \3\s\in \5S(\bh y).
\end{align*}

We now  \sl define  the point source at $i\3y$ \rm by
\begin{align}\lab{dn}\bbx{ \ 
\2\d_n(\3x-i\3y)=-\D_n  G_n(\3x-i\3y),\qq
G_n(\3z)\=\frac1{\o_n}\, \frac{\2r^{2-n}}{n-2} \ }
\end{align}
where $\D_n$ is  the (distributional) Laplacian in $\3x$.
It can be shown \ci{K00} that  for any $\3y$,
$\2\d_n(\3x-i\3y)$ is a  \sl compactly supported  distribution \rm in $\3x\in\rr n$ in the sense of Schwartz \ci{T96}.  Although the proof is somewhat involved, the supports are easily found. For \sl even \rm $n\ge 4$, $G_n(\3z)$ is analytic wherever  $\3z^2\ne 0$, in which case $\D_n G_n(\3z)=0$. Hence\footnote{In the Minkowski space $\4R^{n-1, 1}$,  $\3y$ becomes the time axis,  $\5S$ a slice of the light cone $r=t$, $\5D$ a slice of the future cone $r\le t$,  and   \eq{even} and \eq{odd} translate to  \sl Huygens' principle \rm in even spacetime dimensions and lack thereof in odd spacetime dimensions; see \ci{K00}.
}
\begin{align}\lab{even}
\supp_{\3x} \,\2\d_n(\3x-i\3y)=\5S(\3y)   \hbox{  for even } n \ge 4.
\end{align} 
But for \sl odd \rm $n$, $G_n$ inherits a branch cut from $\2r$, where differentiating across the discontinuity contributes to the support of $\1\d_n$. Thus
\begin{align}\lab{odd}
\supp_{\3x} \,\2\d_n(\3x-i\3y)=\5D(\3y)   \hbox{  for odd } n \ge 3.
\end{align} 
(The same holds for $n=2$ since $2\p G_2(\3z)=-\ln\2r$ has a branch cut on the intreval $\5D$.) The distribution $\2\d_3(\3z)$ will be computed later along with its time-dependent version for pulsed beams. 

To illustrate the above, we work out the  case $n=1$ which, although trivial, is indicative as will be seen.  Recalling our notation $\1x$ for the sign of $x$, we have for $z=x-iy\in\4C$
\begin{align*}
\2r=\sr{(x-iy)^2}=\1x(x-iy)=|x|-i\1x y,\qq x\ne 0.
\end{align*}
(Note that this is \sl not \rm simply the distance in $\4C$ between $iy$ and $x$, which would be $\sr{|x-iy|^2}$ and not $\sr{(x-iy)^2}$.)
Since the `unit sphere' in $\4R$ consists of $x=\pm 1$, its `area' is  $\o_1=2$, and \eq{Lap} gives the correct solution in $\4R$:
\begin{align*}
G_1\0x=-\frac{|x|}2\imp \pl_x^2G_1=-\d_1\0x.
\end{align*} 
Therefore
\begin{gather*}
G_1(x-iy)=-\frac{|x|-i\1x y}2\,,\qq
\pl_xG_1(x-iy)=-\frac{\1x}2+iy\d_1\0x,
\end{gather*}
hence
\begin{align}\lab{d1}
\2\d_1(x-iy)=-\pl_x^2G_1(x-iy)=\d_1\0x-iy\,\d_1'\0x
\end{align}
with Fourier transform
\begin{align}\lab{hatd1}
\widehat{\2\d_1}(k, y)&\=\ir dx\  e^{-i kx}\2\d_1(x-iy)=1+ky.
\end{align}

\section{Point sources in complex spacetime}

Following our  `Euclidean strategy,'  we bring time into the picture by complexifying  Euclidean spacetime: 
\begin{gather*}
x\6E=(\3x,-u)\in\rr4,\qq  y\6E= (\3y, t ) \in\rr4\\
z=x\6E-iy\6E=(\3x-i\3y, -u+it) \in\cc4.
\end{gather*}
This can be rewritten as a complex  \sl Minkowski \rm vector
\begin{gather*}
z=x-iy=(\3z,i\t),\qq \3z=\3x-i\3y,\ \ \t=t-iu,\  \   z^2=\3z^2-\t^2,
\end{gather*}
where 
\begin{gather*}
x=(\3x, it), \      y=(\3y, iu)\in\mink
\end{gather*}
are regarded as \sl real \rm Minkowski vectors with pseudonorms
\begin{align*}
x^2=\3x^2-t^2,\  \   y^2=\3y^2-u^2.
\end{align*}
Now consider the fundamental solution for the Laplacian in $\rr 4$ and its holomorphic extension,
\begin{align}
& G_4 (x\6E)=\frac1{4\p^2x\6E^2},\qqq G_4\0z=\frac1{4\p^2z^2}\notag\\
&\D_4 G_4(\3x, u)=(\D_\3x+\pl_u^2)  G_4(\3x, u)=-\d_4(\3x, u). \lab{eucl}
\end{align}
We have already seen $G\0z=-G_4\0z$ in \eq{Gz}, but there it had the above form only in $\5T\6+$, being antisymmetric because of the sign factor $\1u$ in  \eq{Yx}.   This `twist' was the result of applying the AST to a spacetime field and is another example of the pitfalls of starting from the Lorentzian world.  To be precise, $\1u$ gives the orientation of $\5T\6-$ \sl relative to the way $\mink$ is embedded in $\cc4$ \rm 
(recall the one-dimensional case \eq{ast5}),
thus it has no role in the Euclidean world. In fact, we can now see that the restrictions to $\5T\6\pm$ of the extended relativistic field $\Y\0z$ in \eq{Gz} are part of a single holomorphic function,  just as $\1u\, G\0z$ are both part of $G_4\0z$.

Before defining point sources in complex spacetime, or \bf\sl complex event sources\rm,  we must understand how \sl real \rm point sources in $\mink$ fit into this picture. \sl Naively, \rm it might be hoped that the substitution $u\to it$ into \eq{eucl} gives a propagator  for the wave equation,
\begin{gather}\lab{hyp1}
\square G_4(\3x, t)\=(\D_\3x-\pl_t^2) \,G_4(\3x, t)
\, \top{??}{\sim} -\d_{3,1} (\3x,  t).
\end{gather}
We will see that this fails  for the following reasons:

\bull $G_4$ is singular on the light cone $x^2=0$ and must be defined as a \sl  distribution \rm  in $\mink$ by a limiting process.

\bull When properly defined  in $\mink$, it turns out to be 
 \sl sourceless. \rm

\bull Propagators are related to  \sl causality, \rm which depends on the existence of the light cone and hence  makes sense in $\4R^{3,1}$ but not in $\rr4$. Thus $G_4\0z$, which comes from $\rr4$, cannot itself be a propagator.

To find the extended propagators, note that
\begin{align}\lab{factor}
z^2&=\3z^2-\t^2=\2r^2-\t^2=(\2r-\t)(\2r+\t),
\end{align}
which gives the partial-fractions decomposition already encountered in \eq{split},
\begin{gather}  \lab{prop}
\bbx{\ G_4\0z=i\2D\9-\0z -i\2D\9+\0z,\qq
\2D\9\pm\0z=\frac1{8i\p^2\2r(\t\mp\2r)}\,.\ } 
\end{gather}
Since formal differentiation gives
\begin{align*}
\square \2D\9\pm\0z=0,
\end{align*}
$ \2D\9\pm$ can have sources  only at $\3x\in\5D$  or  $\2r=\pm\t$.  We will show that $\2D\9\pm\0z$  are the proper extensions of the advanced and retarded propagators in $\mink$. Note that choosing the `unphysical' branch of $\2r$ simply reverses the sense of causality:
\begin{align}\lab{negDist}
\2r\to-\2r\imp \2D\9\pm\0z\to -\2D\9\mp\0z.
\end{align}
Thus it will suffice to study the properties of $\2D\9+\0z$.
Its time behavior is governed by the \sl retarded Cauchy kernel: \rm
\begin{align*}
C(\t-\2r)=\frac  1{2\p i(\t-\2r)}
=\frac 1{2\p}\cdot\frac1{(u- q)+i(t- p)},
\end{align*}
which gives the  \sl time-domain radiation pattern \rm  \ci{HY99}. This shows that the ellipsoids $E_p$ are wave fronts, \ie \sl surfaces of constant retardation \rm for $\2D\9+$.   An observer fixed at $\3x$ will see a  pulse peaking at time  $t(\3x)= p(\3x)$, with \sl duration \rm
\begin{align*}
T(\3x)=|u- q(\3x)|.
\end{align*}
The \sl peak magnitude \rm
\begin{align*}
|C(\t-\2r)|_{t= p}=\frac1{2\p|u- q|}
\end{align*}
is constant  along the hyperboloids $H\9\pm_q$. Thus, apart from the attenuation factor $1/\2r$, the peak value of $|\2D\9+\0z|$ remains constant along  the hyperboloids. This shows that  \sl the radiation flows along $H\9+_q$ if $0<q\le a$ \sl and  \rm $H\9-_q$ if $-a\le q<0$.  In the far zone, $C(\t-\2r)$ has peak magnitude
\begin{align}\lab{rad}
\5R\0\q\=|C(\t-\2r)|_{t= r}=\frac1{2\p}\frac1{|u- a\cos\q|}, \qq r\gg a,
\end{align}
showing that there are three qualitatively different cases, depending on the \sl causal \rm character of the imaginary source point.

\begin{enumerate}
  \item  If  $y$ is \sl timelike \rm ($y^2<0$), then $|u|>a$ and $\2D\9+$ is a smooth \sl  pulse \rm outside the source region $\5D$.  Furthermore, the peak radiation pattern $\5R\0\q$ is an  \sl ellipse \rm with \sl  eccentricity $a/|u|$ and  the source at one of the foci. \rm   If $y$ is in the  future cone ($ u>a$), then the semimajor axis of the ellipse points in the direction $\3y$, so that the \sl exploding \rm wave $\2D\9+\0z$ is emitted along $\3y$. It will be shown that $G_4=i\2D\9--i\2D\9+$ is sourceless, so   $\2D\9-\0z$ is an  \sl imploding \rm wave that is simultaneously absorbed along $-\3y$.  

If $y$ is in the past cone ($u<-a$), then  $\2D\9-\0z$ is absorbed along $\3y$ while  $\2D\9+\0z$ is emitted along $-\3y$.  As $y^2\to 0\9-$, the ellipses become more and more eccentric and the  pulsed beams become sharper and sharper. Notice that they have no  \sl sidelobes,\rm\footnote{Sidelobes are angular patterns resulting from interference between waves coming from opposite ends of the source. They can cause problems in radar and communications. 
} 
hence could be useful in applications such as radar tracking and directed communications \ci{K96, K97, K01}.

 \item If $y$ is \sl future lightlike \rm  ($u=a$), then  $\2D\9+$ is  singular  on the \sl ray \rm  along $\3y$  at $t=r$. If $y$ is  \sl past lightlike \rm  ($u=-a$),  $\2D\9+$ is  singular  along $-\3y$ at $t= r$. In either case, the peak radiation pattern $\5R\0\q$ is  \sl parabolic. \rm

  \item  If   $y$ is \sl spacelike \rm ($y^2>0$), then $|u|< a$ and $\2D\9+$ is  singular  at $t= r$ on the  cone  $\cos\q= u/a$, with  a \sl hyperbolic \rm radiation pattern.
 
\end{enumerate}

Only  Case 1 gives reasonable pulsed beams with a chance to be realized with finite energy,  although Cases 2 and 3 should also be of interest since $G_4\0z$ is holomorphic for \sl all \rm $z^2\ne 0$. (Quantum field theory comes to mind, where virtual energy exchange can occur along spacelike directions.)  We therefore assume from now on  that $y\in V\6\pm$, so that $z$ belongs to the \sl causal tube, \rm 
\begin{gather}\lab{T}
z\in\5T=\5T\6+\cup\5T\6-\,,
\end{gather}
as already introduce in \eq{Tpm} starting from Minkowski space.

\sv2
\cl{\includegraphics[width=3 in]{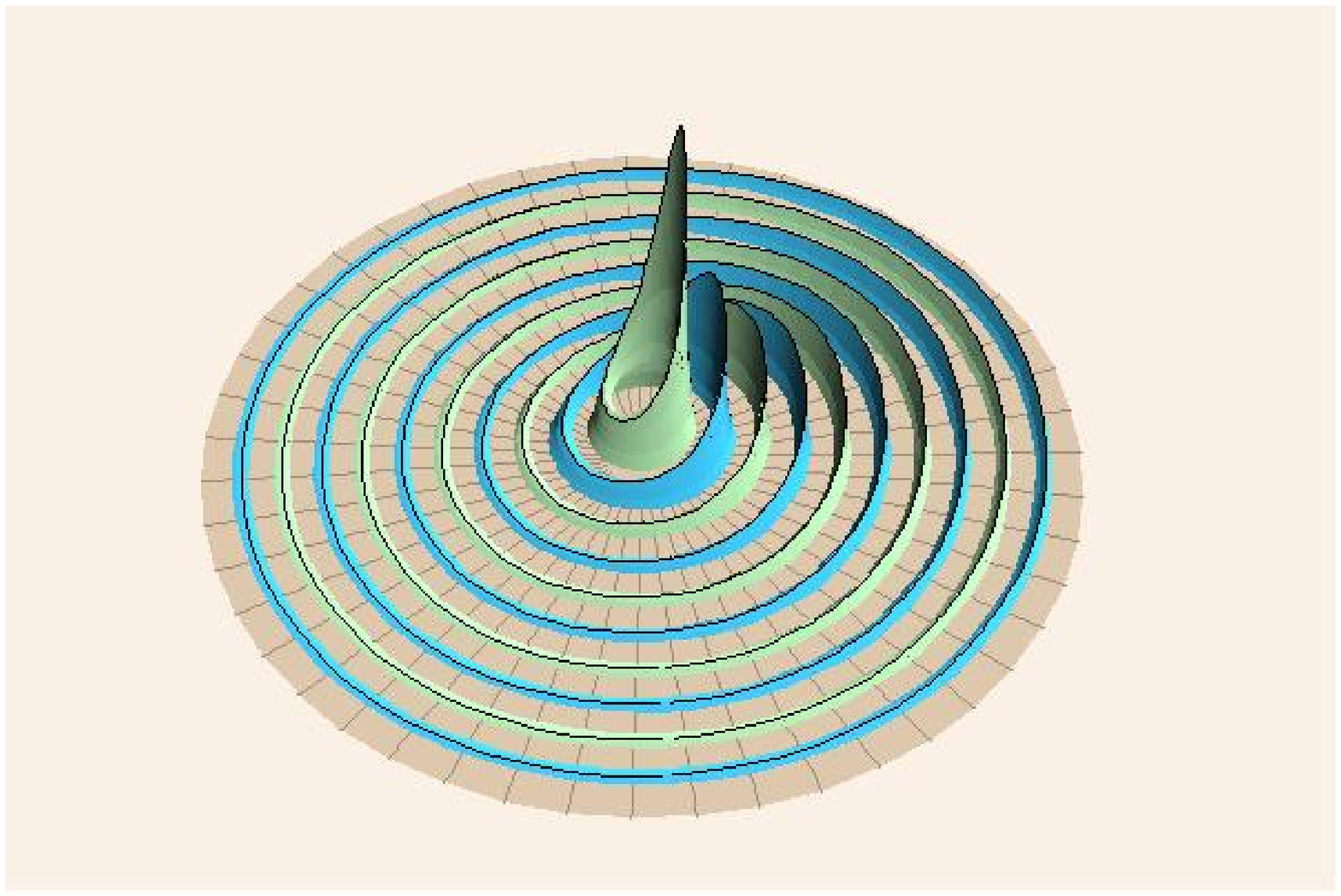}
\includegraphics[width= 3 in]{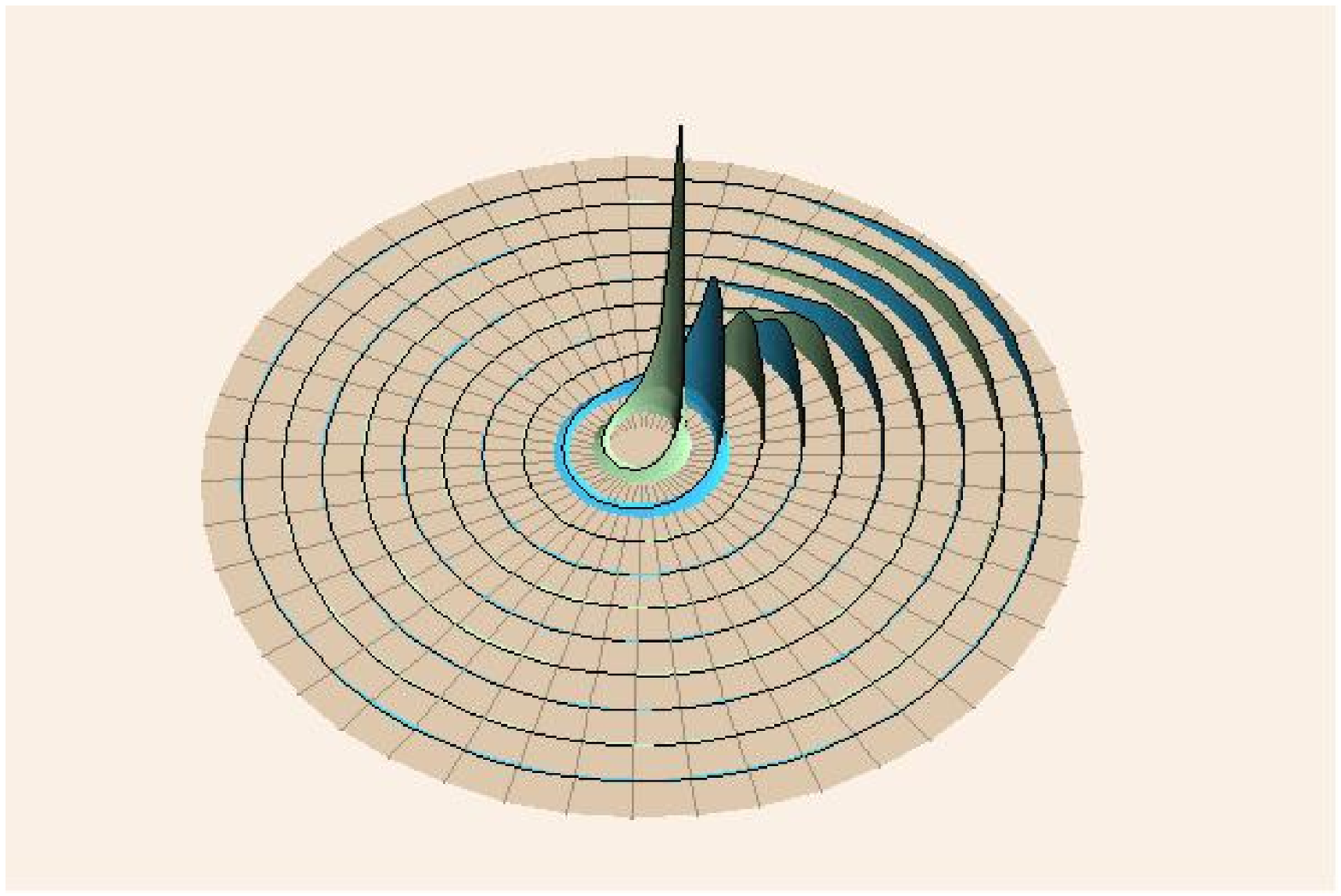}}
\nopagebreak[4]
\cl{\includegraphics[width= 3 in]{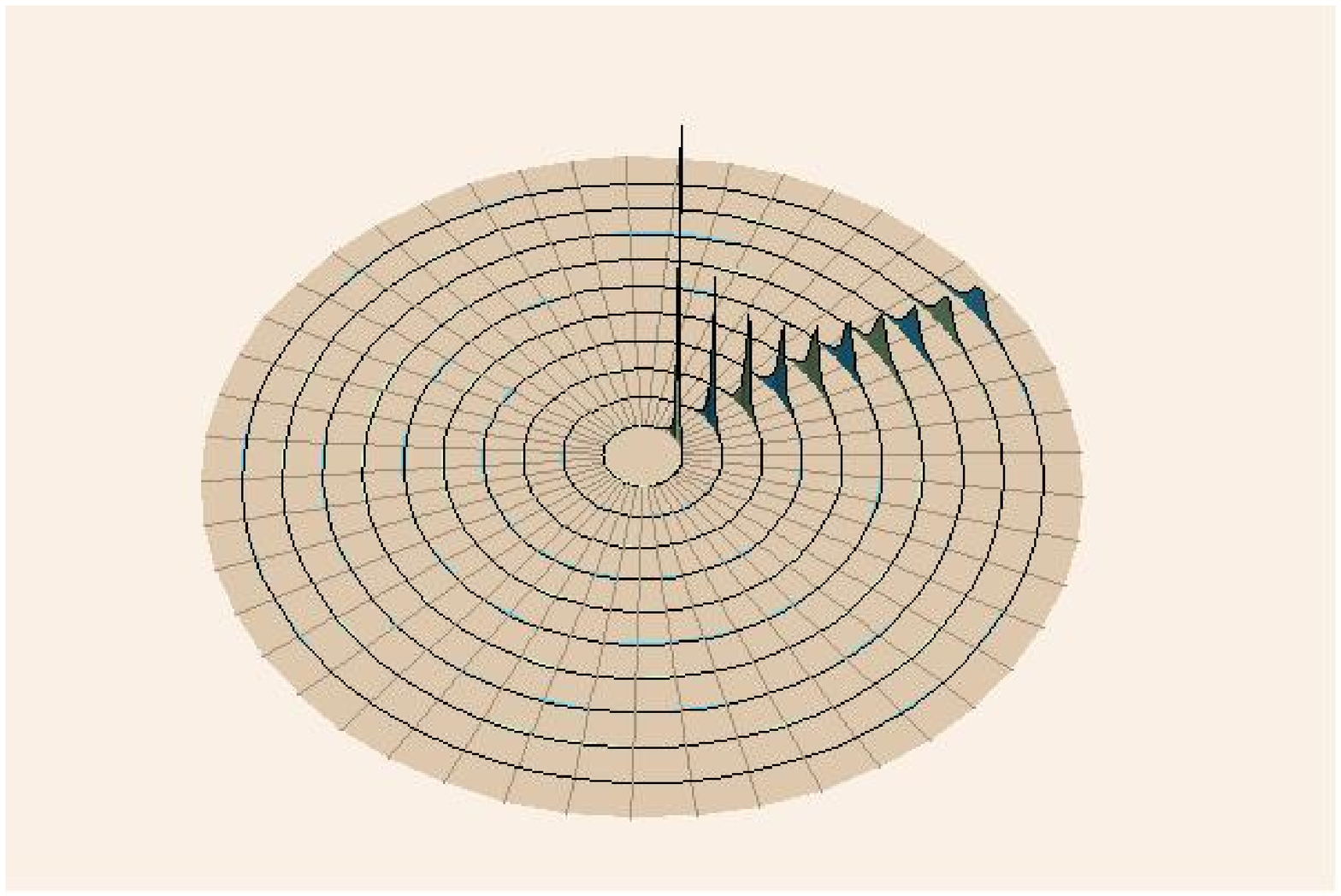}
\includegraphics[width= 3 in]{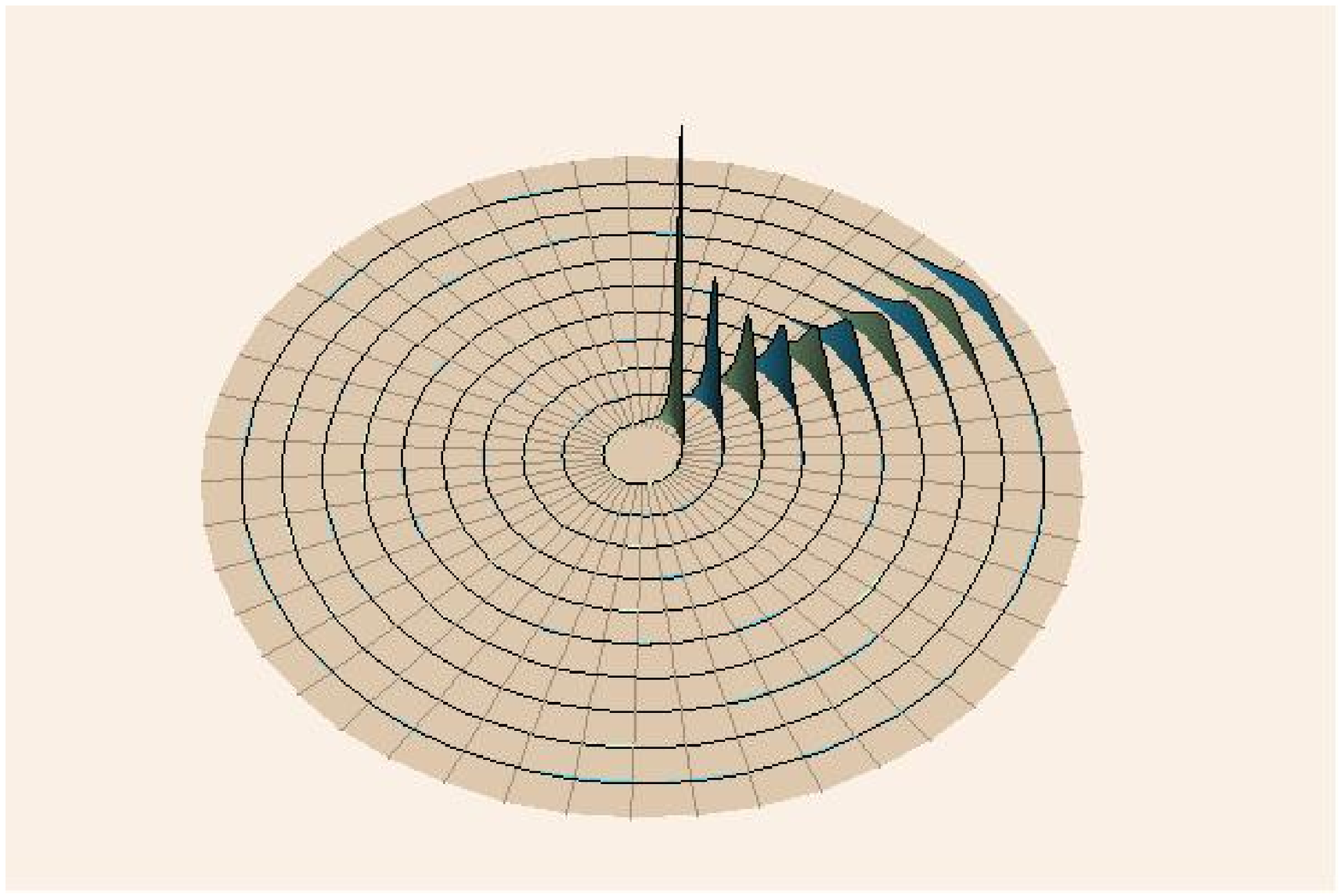}}
\nopagebreak[4]
{\small \bf Figure 2. \rm Time-lapse plots of $|\2D\9+(x-iy)|$ in the far zone, showing the evolution of a \sl single pulse \rm with  propagation vector  $y=(0,0, 1, i u)$. We have taken the slice $x_2=0$, so that the source disk becomes the interval $[-1,1]$ on the $x_1$-axis and the pulse propgates in the $x_3$ direction of the $x_1$-$x_3$ plane. \sl Clockwise from upper left: \rm  $u=1.5,\ 1.1,\ 1.01,\  1.001$.  As $u\to 1$,  $y$ approaches the light cone and the pulsed beams become more and more focused.}
\sv2

We are now ready to learn how $\2D\9\pm\0z$ are related to the wave propagators  in $\mink$. It suffices to focus on $\2D\9+\0z$.  Fix  $y\in V\6+$ and define the partial boundary values
\begin{align}\lab{Gpmx}
\2D\9+(x\pm i 0)&=\lim_{\e\sea 0}\2D\9+(x\pm i\e y).
\end{align}
The limits in \eq{Gpmx} do not depend on the choice of $y\in V\6+$, hence we may choose $y=(\30, iu)$ with $u>0$ so that
\begin{align*}
\2D\9+\0z= \2D\9+(\3x, \t)&=\frac 1{8i\p^2 r}\cdot\frac1{\t-r},
\qq \t=t-iu,
\end{align*}
and the Plemelj jump conditions \ci{T96} give the distributional relations
\begin{align*}
 \2D\9+(x\mp i0)
&=\frac 1{8i\p^2 r}\lim_{\e\sea 0}\frac1{t-r\mp i\e u}
=\frac 1{8i\p^2 r}\5P\frac1{t-r}
\pm \frac {\d(t-r)}{8\p r}
\end{align*}
where $\5P$ is the Cauchy principal value. If we require \sl Huygens' principle \rm \ci{BC87} to be valid  in $\mink$, then the principal value terms must be eliminated and the only combination acceptable as retarded propagator  is
\begin{align}\lab{Minklim}
 D\9+\0x\= \2D\9+(x-i0)- \2D\9+(x+i0)
=\frac{\d(t-r)}{4\p r}\,.
\end{align}
This combination of boundary values, already encountered earlier, will 
 be called the  \sl Minkows\-kian limit. \rm  It does indeed give
\begin{align}\lab{Minkprop}
\square  D\9+\0x=-\d_3(\3x)\d\0t=-\d_{3,1}\0x.
\end{align}

\sv2
\cl{\includegraphics[width=6.5 in]{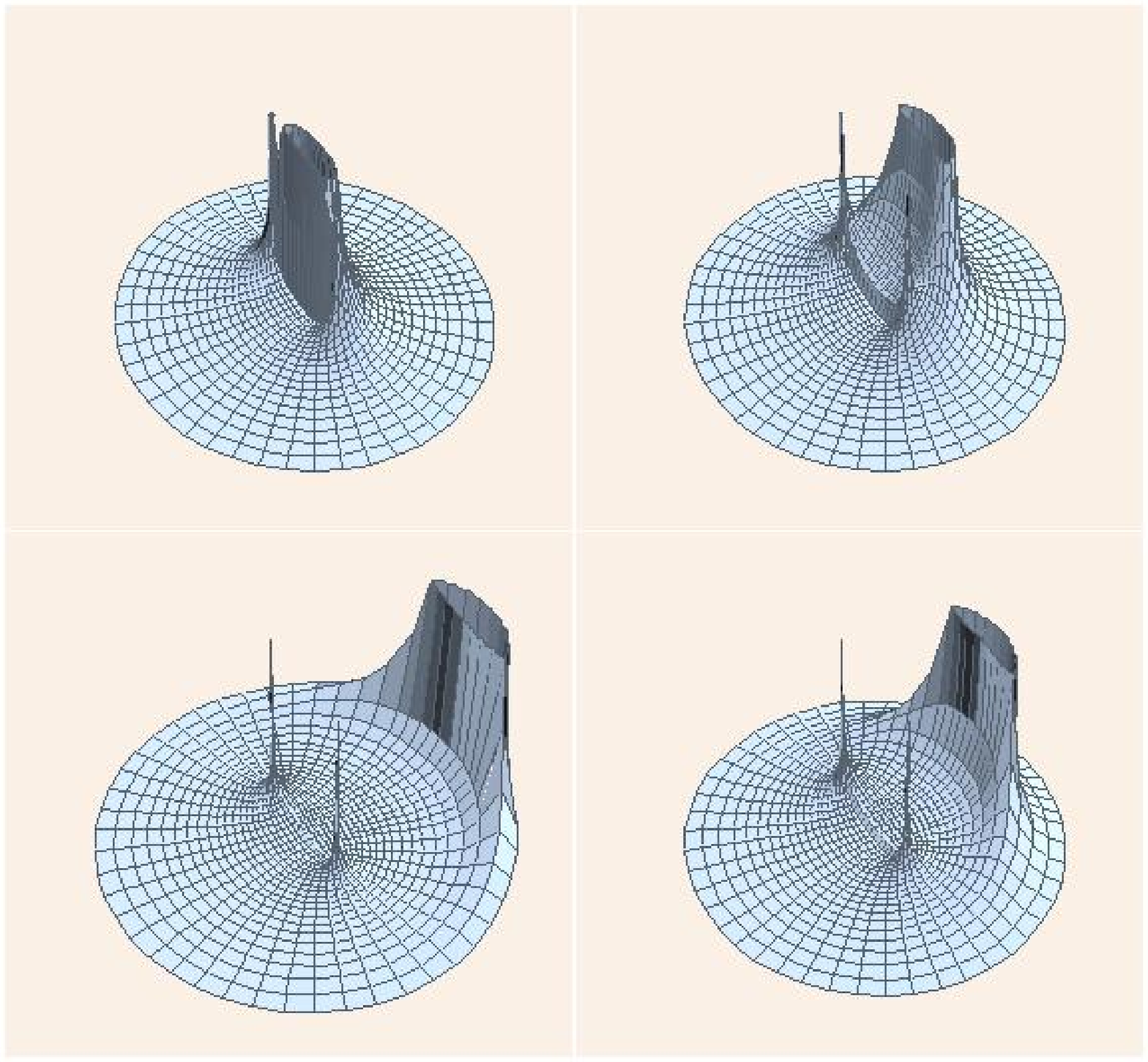}}
\nopagebreak[4]
{\small \bf Figure 3. \rm $|\2D\9+(x-iy)|^2$ with $y=(0,0, 1, 1.01 i)$ in the near zone immediately after launch,  evolving in the $x_1$-$x_3$ plane with $x_2=0$ as in Figure 2.  \sl Clockwise from upper left: \rm
$t = 0.1,  1,  2 , 3.$ The ellipsoidal wave fronts and hyperbolic flow lines are clearly visible. The top of the peak is cut off to show the behavior near the base. The spikes represent the branch circle, whose slice with $x_2=0$ consists of the points $(\pm 1,0,0)$.}
\sv2

\normalsize

The substitution $r\to -r$ as in \eq{negDist} now gives \begin{gather*}
 D\9-\0x\= \2D\9-(x-i0)- \2D\9-(x+i0)=\frac{\d(t+r)}{4\p r}\\
\square  D\9-\0x=-\d_{3,1}\0x.
\end{gather*}
Equation \eq{Minkprop} is the desired hyperbolic counterpart of \eq{eucl}. Its derivation confirms the points made about the failed attempt \eq{hyp1}: 

\bull The distributional  limit \eq{Minklim} played a key role in reproducing Huygens' principle.

\bull By \eq{prop}, the Minkowskian limit of $iG_4\0z$ is
\begin{align} 
R\0x&\=iG_4(x-i0)-iG_4(x+i0)
= D\9+\0x- D\9-\0x \notag\\
&=\frac {\d(t-r)}{4\p r}-\frac {\d(t+r)}{4\p r}\,. \lab{riem}
\end{align}
This is the \sl Riemann function \rm \ci{T96}, which solves the following  initial-value problem of the \sl sourceless \rm wave equation:
\begin{gather}\lab{riemann}
\square R\0x=0,\ \  R(\3x,0)=0,\   \  \pl_t R(\3x,0)=\d_3(\3x). 
\end{gather} 
$iG_4\0z$ can therefore be considered  the \sl extended \rm Riemann function.  It is also identical with Synge's \sl elementary wave function \rm \ci{S65}.

\bull Causality \rm has no meaning for $G_4\0z$ and appears only when  a branch cut is chosen for $\2r$, as  Equations \eq{prop} and \eq{negDist} confirm.

The limits \eq{Minklim} and \eq{riem} are typical of \sl hyperfunction theory \rm  \ci{K88, I92},  where distributions are represented as differences of boundary values of functions  holomorphic in `local' wedge-like domains surrounding the support. In general, there is no preferred  set of such domains and it is necessary to use sheaf cohomology, which makes the theory rather abstract. In our case, however, the two domains $\5T\6\pm$ suffice due to the natural cone structure of relativistic equations.

Following \eq{Minkprop}, we now define  the \sl  point source at $iy$ \rm as
\begin{align}\lab{d31}\bbx{\ 
\2\d_{3,1}(x-iy)=-\square_x  \2D\9\pm(x-iy),\  \ y^2<0\,.   \  }
\end{align}
We list some of its basic properties.

\bull \sl The left side of \eq{d31} is independent of the sign on the 
right. \rm  
Recall that $\square G_4\0z=0$ wherever $G_4$ is holomorphic.
For $z=x-iy\in\5T$,
\begin{align*}
z^2=0\imp x^2=y^2<0 \hbox{ and } x\cdot y=0,
\end{align*}
but $x\cdot y$ cannot vanish since both vectors are timelike. Hence $G_4\0z$ is holomorphic in $\5T$ and
\begin{align*}
\square\,  \2D\9+\0z -\square\,  \2D\9-\0z=\square\,i G_4\0z=0.
\end{align*}

\bull \sl The Minkowskian limit of $\2\d_{3,1}\0z$ is $\d_{3,1}\0x$. \rm  By definition,
\begin{align}
\2\d_{3,1}\0x&=\2\d_{3,1}(x-i0)-\2\d_{3,1}(x+i0)\notag\\
&=-\square  \2D\9+(x-i0)+\square  \2D\9+(x+i0)\notag\\
&=-\square  \2D\9+\0x=-\d_{3,1}\0x.\lab{d31 lim}
\end{align}
\bull \sl  $\2\d_{3,1}(x-iy)$ is supported in the world tube  swept out in $\mink$ by $\5D(\3y)$ at rest: \rm
\begin{align*}
\supp \2\d_{3,1}(x-iy)=\{(\3x, it): \3x\in\5D(\3y)\}\=\tc D(\3y),
 \qq \forall y\in V\6\pm\,.
\end{align*}
This  follows since $\square  \2D\9+=0$ outside the singularities and
\begin{align*}
-\im(\t-\2r)=u-q\ne 0\qq\forall z\in\5T,
\end{align*}
so the only singularities come from the $1/\2r$ factor.

Note that while the Minkowskian limits $ D\9\pm\0x$ and $\d_{3,1}\0x$ are Lorentz-invariant, the extended propagators $ \2D\9\pm\0z$ and their sources $\2\d_{3,1}\0z$ are \sl frame-dependent, \rm the preferred frame being the rest frame of $\5D$.  This can be traced back to the fact that we have obtained the factorization \eq{factor} and associated  splitting \eq{prop} by choosing a branch cut for $\2r(\3z)$ in a particular Lorentz frame. Of course, since $G_4\0z$ is Lorentz-invariant, we may choose to do the splitting in any other frame.

\section{Driven complex  sources}

Suppose we `drive'  a point source fixed at $iy$ with a real time signal  $g_0\0t$. The resulting retarded wave is the convolution \begin{align}
W(\3z,  \t)&=\ir dt' \  \2D\9+(\3z, \t-t')g_0(t')  \notag\\
&=\frac 1{8i\p^2\2r}\ir \frac{g_0(t') dt'}{\t-t'-\2r}  \notag\\
&=\frac{g(\t-\2r)}{4\p\2r}=-g(\t-\2r)\, G_3(\3z),  \lab{Wz}
\end{align}
where $G_3(\3z)$ is the \sl holomorphic Coulomb potential \rm \ci{N73, K01a} and
\begin{align}\lab{ast6}
g\0\t\=\frac1{2\p i}\ir \frac{g_0(t') dt'}{\t-t'}
\end{align}
is the analytic-signal transform \eq{ast4}  of $g_0$. 
The source distribution for the associated field 
\begin{align}\lab{Wz1}
W(\3z, \t)=\frac{g(\t-\2r)}{4\p\2r}
\end{align}
is defined by
\begin{align}\lab{Sz}
S\0z=-\square_x W\0z.
\end{align}
Since $g$ is analytic off the real axis and
\begin{align*}
-\im(\t-\2r)=u-q\ne 0\qq\forall z\in\5T,
\end{align*}
it follows that $W\0z$, like $ \2D\9\pm\0z$, is analytic outside the world tube $\tc D$ swept out by $\5D(\3y)$ at rest. Moreover, formal differentiation gives
\begin{align*}
\square W\0z=0,
\end{align*}
therefore $S(x-iy)$ is also supported in $x\in\tc D$.

\bf\sl Examples: \rm We give three driving signals that will be needed later, with their ASTs and radiated waves \eq{Wz1}:
\begin{align}
g_0\0t&=\d(t) && g(\t)=\frac 1 {2\p i\t}&& W= \2D\9+\0z \lab{as1}\\
g_0\0t&\=1 &&  g(t-iu)=\1u/2 && W=-(\1u/2)\,G_3(\3z)  \lab{as2} \\
g_0\0t&=e^{-i\o t} && g\0\t=\1C(\o, u)\,e^{-i\o t} && W\0z=
g\0\t\,B_\o(\3z), \lab{as3}
\end{align}
where $\1C$ is the Fourier transform of the Cauchy kernel \eq{hatC}
\begin{align}\lab{Cauchy2}
\1C(\o, u)&=\1u\,\Q(\o u)\,e^{-\o u}
\end{align}
and
\begin{align}\lab{beam}
B_\o(\3z)=\frac{e^{i\o\2r}}{4\p\2r}
\end{align}
is the \sl time-harmonic complex-source beam, \rm applied widely in engineering \ci{HF01}. In the last equality of \eq{as3} we used 
\begin{align}\lab{sign}
|q|\le a\imp \sgn(u-q)=\sgn u\imp\Q(\o(u-q))=\Q(\o u).
\end{align}
Note that to make \eq{as2} a special case of \eq{as1} with $\o=0$, we must define $\Q\00=1/2$.

The far-zone approximation \eq{far} to \eq{as3}
shows that $W$ is indeed a beam in the direction of $\1u\,\3y$ which becomes more and more focused as $y$ approaches the light cone:
\begin{align} \lab{thBeam}
r\gg a\imp W\0z\app
 \1u\,\Q(\o u)\,e^{-\o(u-a\cos\q)}\,\frac{e^{-i\o(t-r)}}{4\p r}\,.
\end{align}
This is not surprising, since \eq{as3} is a Fourier component of the pulsed beam \eq{as1}.  The beam $W$ is exponentially stronger for $\1q=\1u$ than for $\1q=-\1u$, so the beams are \sl directed. \rm Equation \eq{thBeam} shows that \sl the  imaginary retardation  
$u\to u-q$ serves to focus the beam. \rm 

Just as the spatial displacement of a point source from $\3x=\30$  to
$i\3y$ expands it to a disk, the temporal displacement of the impulse  from $t=0$ to $iu$ gives it duration, as seen in \eq{as1}. We may interpret the parameter $|u|$  as a \sl response time \rm for the source at $iy$. A large  response time suppresses rapid variations in the driving signal. This is  modeled\footnote{Admittedly this model leaves something to be desired since $g$ decays slowly even if $g_0$ has compact support. In particular, it is not even causal: the vanishing of $g_0\0t$  for $t<0$ does not imply the same for $g(t-iu)$; in fact, the latter cannot vanish on \sl any \rm interval because it is analytic.  One way to improve the model is to represent the response by derivatives of $g\0\t$,  as in \eq{split}. This suppresses the low frequencies and amounts to putting $g_0$ through a `band-pass filter' of the form  
$\o^\a\Q(\o u) e^{-\o u}$centered around $\o=\a/u$.}  
by the factor $\Q(\o u)\,e^{-\o u}$, which acts as a \sl filter \rm to smooth the signal. The larger the source, the longer the response time since the excitation can travel to different parts of the source before emitting a wave. Thus, a rough but intuitive way of understanding the timelike character of $y$ is to note that the time needed  for a signal to travel from the center to the rim of the disk is $a$, and therefore $|u|>a$.

With the interpretation of $u$ as a `response time,'  all four components of $y$ have a direct significance in terms of the \sl source \rm itself,  without reference to the radiated beam.

\section{Main results on sources}\tlab{S:results}

In this section we present our main results on the source distribution for 
beams $W\0z$ of the general type \eq{Wz1}. These will be proved rigorously in the Appendix.  The choices \eq{as1}--\eq{as3} for $g_0$ yield the sources for pulsed, static, and time-harmonic beams generated by \sl complex event \rm sources. Furthermore, choosing a plane wave for the test function will yield the spacetime Fourier transforms of the sources and their beams, giving valuable insight into their propagation properties and making them a potentially useful computational tool.

For clarity, the results are stated as theorems. However, I have included some discussion to help make them digestible for readers without training in the art of arid mathematical discourse.  A more extensive discussion of the results and their interpretation is given in the next  section.

To compute $S\0z$ as a distribution, we must deal with its singularities. This will be done by \sl shielding \rm $\5D$  with an ellipsoid $E_{\e a}$ and taking the limit $\e\to 0$ once the computations are complete. Thus let $\e>0$ and
\begin{align*}
W_\e\0z=\Q(p-\e a) W\0z
=\begin{cases}
    W\0z  & \text{if $\3x$ is outside } E_{\e a}\\
    0  & \text{if $\3x$ is inside } E_{\e a},.
\end{cases}
\end{align*}
The singularities on $\5D$, consisting of the discontinuity in the interior of the disk and the divergence on the boundary $\5S=\pl\5D$, have been replaced by a \sl uniformly finite \rm jump discontinuity across $E_{\e a}$.  The  \sl regularized source,  \rm defined by
\begin{align*}
S_\e\0z\=-\square W_\e\0z,
\end{align*} 
is therefore  supported on the world tube swept out by $E_{\e a}$ at rest,
\begin{align*}
\2E_{\e a}=\{x=(\3x, it): \3x\in E_{\e a}\}.
\end{align*}
The `bare' source $S$ will then be defined by
\begin{align}\lab{weakl}
S\0z=\lim_{\e\sea 0} S_\e\0z,
\end{align}
where the limit is taken in the  distributional (weak) sense by `smearing' over test functions.

\bf\sl Note: \rm The definition \eq{weakl} is necessary on \sl conceptual \rm as well as  technical grounds for the following subtle reason. We have chosen a branch of the complex distance  $\2r$ to reduce to the usual positive distance as $\3y\to\30$, but nowhere have we actually \sl enforced \rm this in our equations --- until now!  The definition  \eq{weakl} clearly communicates our choice to the equations.

\begin{thm}\tlab{T:shielded}
{\bf (Shielded complex source)}

For given timelike $y$, $S_\e(x-iy)$ is a (Schwartz) distribution  
{\rm \ci{T96}} in $x\in\mink$ supported on $\2E_{\e a}$. It is regular in $t$ (no smearing necessary) and acts on a spatial test function 
$f(\3x)$ as follows:
\begin{gather}\lab{Sef}\bbx{\ 
\la S_\e\,, f\ra=\frac{\a^*\a}{2a}\lb \frac{g\`f}{i\2r}\rb_{\2r=\a^*}^{\2r=\a}
+\frac{\a^*\a}{a}\int_{-a}^a \frac{dq}{\2r}\  g \`f_\2r\Bigm|_{p=\e a}\ }
\end{gather}
where $\a=\e a- ia$ and $\a^*$ are the {\rm north and south poles} of $E_{\e a}$ and we used the notation
\begin{align}\lab{mean}
\`f(\2r)&\=\`f(p,q)=\frac1{2\p}\int_0^{2\p} f(p, q,\f) d\f\\
2 \`f_\2r&=2\pl_\2r \`f=\pl_p\`f+i\pl_q\`f \notag
\end{align}
without implying analyticity in $\2r=p-iq$. $S_\e$ is given directly (without smearing) as
\begin{gather}\lab{Se}\bbx{\ 
S_\e\0z=W\0z \LB i\d(\2r-\a) - i\d(\2r-\a^*) 
-\Bigm| \frac{\a}{\2r}\Bigm|^2\d(p-\e a)\,(\pl_p+i\pl_q) \RB  \ }
\end{gather}
where
\begin{align*}
\d(\2r-\a)=\d(p-\e a)\d(q- a),\qq \d(\2r-\a^*)=\d(p-\e a)\d(q+a)
\end{align*}
are {\rm point sources} at the north and south poles of $E_{\e a}$.
The  normal derivative $\pl_p$ is interpreted as a {\rm double layer}, and the  tangential derivative $\pl_q$ as a  {\rm surface flow}  on $E_{\e a}$. The entire distribution is modulated in space and time by the values of the beam $W\0z$ on $E_{\e a}$, including the retarded analytic signal $g(\t-\2r)$ of its driving function responsible for collimating the beam. The distribution {\rm\eq{Se}} must be applied as a differential operator to a test function, then integrated.

\end{thm}

\begin{thm}\tlab{T:bare}
{\bf (Unshielded complex source)}

For given timelike $y$, $S(x-iy)$ is a distribution in $x\in\mink$,
regular in $t$,  acting on a spatial test function $f(\3x)$ by
\begin{gather}\lab{Sfq}
\la S\,, f\ra=\2g(\t, a)f(\30)
+ia\int_0^a \frac{dq}{q}\  \2g(\t,q)\lb \`f_p(iq)+i\`f_q(iq) \rb,
\end{gather}
where
\begin{gather}\lab{gtilde}
\2g(\t, q)\=\frac12\lb  g(\t+iq)+  g(\t-iq)\rb
\end{gather}
is the average of $g(\t-\2r)$ over the jump across  $\5D$.
The integral  converges because the continuity of $f(\3x)$ across $\5D$ and its differentiability on $\5D$ imply that the partial derivatives of $\`f$ on $\5D$ are $\5O\0q$:
\begin{gather}\lab{fpfq}
\`f_p(iq)=\frac q a\, \`f_\x\,,\qq \`f_q(iq)=-\frac q \r \, \`f_\r\
\end{gather}
where $\`f_\r=\pl_\r\`f,\  \`f_\x=\pl_\x\`f$.  In cylindrical coordinates,
with $q\=\sr{a^2-\r^2}$, 
\begin{gather}\lab{Sf}\bbx{\ 
\la S\,, f\ra=\2g(\t, a)f(\30)+\int_0^a \frac{d\r}{q}\ 
\2g(\t, q)\lb a \`f_\r+i\r\`f_\x \rb.\ }
\end{gather}
$S$ can be expressed without smearing as
\begin{gather}\lab{S}\bbx{\ 
S\0z=\2g(\t,q) \LB \d_3(\3x)
+\frac{\Q(a-\r)\d\0\x}{2\p q\,\r}
\lp a \pl_\r+i\r\pl_\x \rp\RB.\ }
\end{gather}
Therefore, $S\0z$ has  a {\rm real point source} at the origin and a  surface distribution on $\5D$ consisting of a {\rm radial flow} $\pl_\r\,$  and a {\rm double layer}  $\pl_\x$. The entire distribution is modulated in space and time by $\2g(\t,q)$.

\end{thm}

\bf\sl Remark 1: \rm  Constructions of complex sources exist in higher dimensions \ci{K0x}, but  \sl only for  $n\le 3$  does $\2\d_n\0z$ contain a real point source. \rm  (See Equation \eq{d1} for the trivial case $n=1$.) The reason is that  for $n>3$, $S$ inherits the factor  $\r^{n-3}$ from the volume element, and this vanishes at the poles of $E_{\e a}$.

\bf\sl Remark 2: \rm The point source $\d_3(\3x)$ in \eq{S} is not apparent in Figure 3, which seems puzzling. However, recall that this point source is the limit of the two point sources in  \eq{Se} and,  as explained below \eq{Sef2}, these point sources are `removable singularities' resulting from an integration by parts. Without them, $S_\e\0z$ has the equivalent `local' expression \eq{Se3}. However, that expression does not have a limit as $\e\to 0$ and must be \sl regularized \rm by subtracting and adding a Taylor approximation to  the numerator, as was done in \ci{K00} for the general case of $\rr n$. (This procedure gives a generalization of the Cauchy principal value to arbitrary order; see \ci{GS64}.) The resulting limit distribution $\la S, f\ra$ is \sl nonlocal, \rm  \ie it cannot be written in unsmeared form like  \eq{S}.   We write `local' is in quotes because  \eq{S} still contains differential operators. However, these can be interpreted locally as double layers and flows, which would be impossible if $\la S, f\ra$ had subtractions. This is especially important if one is interested in the possibility of \sl realizing \rm such sources, \ie building instruments that can emit and, by reciprocity,  detect pulsed beams of the above type.    

\bf\sl Remark 3: \rm An even more intriguing possibility is that some naturally occurring  fundamental processes can be associated with emission, absorption, or radiation  by complex source points. Natural extensions of the present construction exist for Maxwell's equations \ci{K02}, and even for massive Klein-Gordon and Dirac fields \ci{K90}, although the sources in the latter cases have not yet been studied. In fact, one of the first applications of what I have called complex distance occurred in the derivation by Newman et al. of charged spinning black holes; see \ci{N65, N73}, and the recent work \ci{K01a, N02}.

\sv2

\begin{cor}\tlab{T:cases}
{\bf (Event sources, static and time-harmonic point sources)}

Letting $g_0\0t=\d\0t$  as in {\rm\eq{as1}} gives $g\0\t=1/2\p i\t$ and
\begin{align*}
i\2g(\t,q)=-\frac \t{2\p(\t^2+a^2-\r^2)}
=\frac \t{2\p z^2}\,,\qq z\in\tc D.
\end{align*}
Therefore the  {\rm event source} at $iy$ is
\begin{align}\lab{2d31}\bbx{\ 
\2\d_{3,1}\0z=\frac \t{2\p z^2}\LB\d_3(\3x)+ 
\frac{\Q(a-\r)\d\0\x}{2\p q\,\r}\lp a \pl_\r+i\r\pl_\x \rp \RB.\ }
\end{align}
Letting $g_0\0t\=1$ as in {\rm\eq{as2}} gives the {\rm static point source} at  $i\3y$:
\begin{align}\lab{d3}\bbx{\ 
\2\d_3(\3z)=\d_3(\3x)+ 
\frac{\Q(a-\r)\d\0\x}{2\p q\,\r}\lp a \pl_\r+i\r\pl_\x \rp. \ }
\end{align}
Letting $g_0\0t=e^{-i\o t}$ as in {\rm\eq{as3}} gives the analytic signal for the {\bf\sl time-harmonic point source at} $i\3y$:
\begin{align}\lab{TH}
\2g(\t,q)
=\1u \,\Q(\o u) e^{-i\o \t}\cosh(\o q).
\end{align}

\end{cor}

Much valuable information is gained by computing the Fourier transforms of the pulsed beams and their sources. Although this seems at first a difficult task  because of the convoluted spatial dependence of the beams through $\2r$ and its oblate spheroidal coordinates, the result is intriguing and  `unreasonably' simple.   

Recall our notation for Minkoswki space  $M$  and its dual Fourier space  $M'$. Given a spatial direction $\bh y$,  we use the cylindrical coordinates \eq{cyl} 
\begin{align}
&x=(\3\r, \x,it)&& k=(\3h, l, i\o),&&\3\r\cdot\bh y= \3h\cdot\bh y=0\notag\\
& \x=\bh y\cdot \3x && \r=|\3\r|=\sr{r^2-\x^2} &&  r=|\3x| \notag\\
& l=\bh y\cdot \3k && h=|\3h|=\sr{\k^2-l^2} && \k=|\3k| \lab{cyl4}
\end{align}
so that $l$ and $h$ are the \sl longitudinal and transverse wave numbers  \rm with respect to  $\bh y$ and the pairing between $M'$ and $M$ is  
\begin{align}\lab{cyl5}
k\cdot x=\3k\cdot\3x-\o t=\3h\cdot\3\r+l \x-\o t=h\r\cos\f+l  \x-\o t .
\end{align}

\begin{thm}\tlab{T:hatSe}
{\bf (Fourier transform of shielded source)}

\bull Given  $y\in V$ with $|\3y|=a>0$ and $\e>0$,  define the  complex wave vector $k_\e$
\begin{align}
& \o_\e =\o-i\e  l    && h_\e=|\h|\,h &&\boxed{\ \h\=\e-i=\a/a\ } \notag\\ 
& l _\e  = l -i \e \o &&  \f_\e=\f ,  \lab{ke}
\end{align}
which satisfies
\begin{align}\lab{ke2 }
k_\e^2=|\h|^2 k^2=k^2+\e^2k^2
\end{align}
and thus preserves the {\bf\sl complex light cone.}
Then the Fourier transform {\rm \eq{hatast}} with respect to $x$ of the shielded source $S_\e(x-iy)$  is
\begin{align}\lab{hatSe}
\bbx{\  \1S_\e(k, y)
=\1g(\o, u)\,e^{i\e\o a}\,\O(k_\e, \3y), \qq
\O(k_\e, \3y)=\cos(\m_\e a)+\frac{  l _\e}{\m_\e}\,\sin(\m_\e a) \   }
\end{align}
where 
\begin{align}\lab{hatg}
\1g(\o, u)=\1u \,\Q(\o u)\,e^{-\o u} \, \1g_0\0\o=\1C(\o, u)\,\1g_0\0\o
\end{align}
and
\begin{align}\lab{me}
\m_\e &=\sr{h_\e ^2-\o_\e ^2}=\sr{k_\e ^2-  l _\e ^2}\,. 
\end{align}
The transformation $k\mt k_\e$ is the product of a  {\bf\sl scaling} $k\mt |\h| k$ and a {\bf\sl real rotation}   in the $  l  $-$i\o$ plane, or {\bf\sl imaginary Lorentz transformation} in the $  l  $-$\o$ plane, given by  
\begin{gather*}
\ l\mt \frac{  l  -\e i\o }{\sr{1+\e^2}},\qqq
i\o\mt \frac{i\o+\e   l  }{\sr{1+\e^2}}\,.
\end{gather*}

\bull  $\O$ plays the role of a {\bf\sl focusing filter} off the light cone, and  on the light cone it simplifies as
\begin{align}\lab{lightcone}
k^2=0 \iff   l =\pm i \m \iff \O(k, y)=e^{\pm i\m a}.
\end{align}
Since $\m$ is imaginary for propagating waves and real for evanescent waves, $\O$ amplifies `forward' waves and dampens `backward' waves. It does not depend on the branch of the square root  since  it is even in $\m $.

\end{thm}

\begin{cor}\tlab{T:hatS}
{\bf (Fourier transforms of bare sources)}

The Fourier transform of the bare source $S(x-iy)$ is 
\begin{align}\lab{hatS}\bbx{\ 
\1S(k, y)=\1g(\o,u)\,\O(k, \3y)
=\1g(\o,u)\LB \cos(\m a)+\frac   l  \m\,\sin(\m a) \RB\ }
\end{align}
where 
\begin{align*}
\m=\sr{h^2-\o^2}=\sr{k^2- l ^2}
\end{align*} 
can be real or imaginary.  The Fourier transform of the {\bf\sl event source}  at  $iy$ is 
\begin{gather} \lab{hatd31}
\bbx{\  \wh{{\2\d}}_{3,1}(k, y)
=\1C(\o,u)\,\O(k, \3y),
\qq \1C(\o,u)=\1u\,\Q(\o u) e^{-\o u} \  }
\end{gather}
and that of the point source at $i\3y$ is 
\begin{align}\lab{hatd3}
\bbx{\  \widehat{\2\d_3}(\3k,\3y)= \cos(h a)+\frac{  l   }{h}\,\sin(h a). \ }
\end{align}

\end{cor}

Amazingly, complex sources are much simpler in Fourier space than in space or in spacetime! This shows that effective computations can be performed with them and their radiated beams using `fast' numerical methods like the FFT.  For example, spacetime convolutions of complex sources  with arbitrary `densities' can be performed with ease by multiplying the Fourier transforms.

The surprising simplicity and symmetry of the above expressions may be a `miracle,' but it cannot be an accident.  Its origin and some consequences are discussed in the next section.

Note that the expression \eq{hatd3} is actually valid for 
$\wh{{\2\d}}_n(\3k, \3y)$ with \sl any \rm value of $n\ge 1$, as will be proved elsewhere. In particular, note that it holds in the trivial case $n=1$ \eq{hatd1}, where $a=y,\ l=k,$ and $h=0$.

\section{Interpretation and discussion of results}\tlab{S:discussion}

Let us attempt to understand some of the expressions given in the last section. Our discussion is necessarily somewhat speculative, undertaken with the desire to add qualitative value to the raw mathematical equations.

\bull \bf\sl The bare source $S\0z$:  \rm Looking at Equation \eq{S},
\begin{align}\lab{S2}
S\0z=\2g(\t,q) \LB \d_3(\3x)
+\frac{\Q(a-\r)\d\0\x}{2\p \,\sr{a^2-\r^2}\,\r}
\lp a \pl_\r+i\r\pl_\x \rp\RB\!,
\end{align}
we note that 
\begin{align*}
G_{2,1}(x_1, x_2,a)\=\frac{\Q(a-\r)}{2\p\,\sr{a^2-\r^2}}
\end{align*}
is the \sl retarded propagator for the wave equation in two space dimensions, with $a=|\3y|$ playing the role of time \rm   \ci{T96}:
\begin{align*}
(\D_2-\pl_a^2)G_{2,1}=-\d(x_1)\d(x_2)\d\0a .
\end{align*}
The origin of this analogy is not difficult to find. By complexifying the distance function
\begin{align*}
r(\3x)\to\2r(\3x-i\3y)=\sr{r^2-a^2-2i \3x\cdot\3y},
\end{align*}  
we have implicitly introduced  $a$ as `time' measured along the axis  defined by the unit vector $\bh y$, in exactly the same way as complexifying the Euclidean time $u$ opened up the light cone and gave rise to the physical time $t$.  In a coordinate system where 
$\bh y=(0,0,1)$, the  branch cut $\5D(\3y)$ is simply a slice of the `future cone'
\begin{align*}
V\6+(\bh y)=\{(x_1, x_2, is): \r\=\sr{x_1^2+x_2^2}< s \}\subset\4R^{2,1}
\end{align*}
at  $s=a$, with $\r$ as the distance function in the orthogonal `space' $\rr2$. The Heaviside function $\Q(a-\r)$ thus merely enforces `causality,' and the factor $\sr{a^2-\r^2}$ in the denominator reflects the absence of Huygens' principle in  $\4R^{2,1}$, where an 'observer' at $(x_1, x_2)$ will `hear' nothing for $s<\r$, then a 'sonic boom' at $s=\r$ with a decaying tail for $s>\r$.

Does this give any insight to the nature of $S\0z$? It suggests looking at $a$ as an \sl evolution parameter. \rm  Think of `morphing' from a point source to a disk source by gradually changing $a$, then \eq{S2} shows how the source flows while evolving.

\sv2

\bull \bf\sl The shielded source  \rm \eq{Se}
\begin{align}\lab{Se2}
S_\e\0z=W\0z \LB i\d(\2r-\a) - i\d(\2r-\a^*) 
-\Bigm| \frac{\a}{\2r}\Bigm|^2\d(p-\e a)\,(\pl_p+i\pl_q) \RB\,.
\end{align}
This expression is extremely simple, reflecting the regularity achieved by replacing the singular disk $\5D$ with the oblate spheroid $E_{\e a}$.
The first two terms are a pair of real point sources  at the north and south poles of $E_{\e a}$,
\begin{align*}
\a=\e a-ia=\h a,\qq \a^*=\h^* a.
\end{align*} 
As already mentioned,  the terms with the normal derivative $\pl_p$ and the  tangential derivative $\pl_q$ may be interpreted as a double layer  and a flow on $E_{\e a}$, although this needs some discussion.  Further insight is gained from the Fourier transform \eq{hatSe},
\begin{align}\lab{hatSe4}
\1S_\e(k,y)=\1g(\o,u)e^{i\e a}\,\O(k_\e, \3y),\qq
\O(k,\3y)=\cos(\m a)+\frac{ l}{\m }\,\sin(\m  a).
\end{align}
In the proof of \eq{hatSe4} in the Appendix, we saved the final details for this discussion because they shed light on the nature of the source, and also because they contain some spectacular cancellations and reveal an amazing hidden structure, namely the complex mapping $k\to k_\e$ of Fourier space associated with the replacement of the disk source $\5D$ by the \sl Huygens source \rm  $S_\e$ that generates the identical field  outside $E_{\e a}$ by  emitting  `secondary wavelets.' 

By  \eq{hatSe2},  \eq{I12} and \eq{I3},
\begin{align}
&\O_\e(k, \3y)= I_0-I_1+i\e I_2+\h^*\h I_3\nt\\
&I_0=\cosh(\o_\e  a)-i\e \sinh(\o_\e  a)\nt\\
&I_1=\cosh(\o_\e  a)-\cos(\m_\e  a)\lab{I123}\\
&I_2=\sinh(\o_\e  a)-(\o_\e  /\m_\e )\sin(\m_\e  a)\nt\\
&I_3=(  l  /\m_\e)\sin(\m_\e  a).
\end{align}
The derivation shows that $I_0$ is due to the point sources at $\a$ and $\a^*$ and the other terms are due to the  the combined double layer and flow. Specifically, $I_1$ and $I_2$ come from the $\r$-derivative and $I_3$ comes from the  $\x$-derivative. The first terms of $I_1$ and $I_2$ \sl entirely cancel the point-source term \rm $I_0$, resulting in
\begin{align*}
\O_\e(k, \3y)=\cos(\m_\e  a)-i\e(\o_\e/\m_\e)\sin(\m_\e  a)
+(\e^2+1)(  l  /\m_\e)\sin(\m_\e  a).
\end{align*}
This cancellation is related to the fact that the point sources in $I_0$  resulted from an integration by parts, which in turn was needed  to get a local expression for $S\0z$ (see the note below \eq{Sef2}).
But
\begin{align*}
(\e^2+1)l -i\e\o_\e=(\e^2+1)l -i\e(\o-i\e l) =l -i\e\o =  l _\e
\end{align*}
which gives the final form \eq{hatSe4}.

The cancellations and simplifications taking place to yield this simple result appear to be  `miraculous.'  This could be merely good fortune or,  more likely, an indication that the Fourier sources and possibly also the unexpected complex mapping $k\to k_\e$, are more `fundamental' than the spacetime beams we started with and should therefore be thoroughly understood.

On the practical side, simplicity in the Fourier domain usually means 
enhanced analytical power and the existence of efficient implementations by 'fast' algorithms. The above Fourier sources offer a promising new tool, modeling processes  of  \sl directed emission and absorption \rm  that could be useful in quantum  as well as classical physics.

As a first step, and an example of working with the Fourier sources, 
we  compute the pulsed beam $W\0z$ from its source.

\section{Weyl representations for complex-source beams}

 How does the simple Fourier  source  $\1S(k,y)$ radiate a pulsed beam as complex (!) as $W\0z$? To learn the answer, we now compute $W$ from $\1S(k,y)$.  In the process, we prove a generalization to \sl beams \rm of Hermann Weyl's  representation  of time-harmonic spherical waves. 

From Equations \eq{Wz1}, \eq{Sz} and \eq{hatS}, recall
our path from $W$ to $\1S$:
 \begin{align}
W(\3z, \t)&=\frac{g(\t-\2r)}{4\p\2r}\nt\\
S(x-iy)&=-\square_x W(x-iy) \lab{WS}\\
\1S(k,y)&\=\int_M dx\ e^{-ik\cdot x}\,S(x-iy)=\1g(\o,u)\, \O(k, \3y)\nt\\
\O(k,\3y)&= \cos(\m a)+\frac   l  \m\,\sin(\m a)\nt
\end{align}
where $g$ is  the AST  of a driving signal $g_0\0t$ obtained by convolution \eq{ast4} with the Cauchy kernel $C\0\t$, so that
\begin{align*}
\1g(\o,u)=\1C(\o, u)\,\1g_0\0\o,\qqq \1C(\o, u)=\1u \Q(\o u)\,e^{-\o u}\,.
\end{align*} 
\sl Formally, \rm \eq{WS} implies
\begin{align} \lab{W3}
W(x-iy)=\int_{M'} \db k\  e^{ik\cdot x}\, \frac{\1S(k,y)}{k^2}\,,
\end{align}
but the right side must be defined since $k^2$ vanishes on the light cone. In spacetime terms, this corresponds to the fact that "initial values" must be specified in order to solve \eq{WS} for $W$. More precisely, since we are dealing with $-\8<t<\8$, we need the behavior of $W$ as $t\to\pm\8$, which describes the \sl causal \rm relation between the source and the solution. This amounts to a choice of contour in Fourier space that avoids the light cone singularities, so that the solution can be computed by residues.

The temporal Fourier transform of $W$ is
\begin{gather*}
\frac{1}{4\p\2r}\ir dt\ e^{i\o t}\, g(t-iu-\2r)=\1g(\o, u)\,B_\o(\3z),
\end{gather*}
where 
\begin{align}
B_\o(\3z)=\frac{e^{i\o\2r}}{4\p\2r}\,, \lab{Wo}
\end{align}
is the time-harmonic complex source beam \eq{beam}, and
 translating the integration contour  by $t\to t+\2r$, or $t\to t-iq$, is justified because $g$ is analytic off the real axis and $|q|\le |a|< |u|$.
Therefore, we need to establish that the function
\begin{align}
U(\3z,\o)&\=\int_\rr3\db \3k\ e^{i\3k\cdot \3x}\cdot\frac{\O(k,\3y)}{k^2}
\notag\\
&=\int h\db h\, \db\f\,\db l \ e^{ih\r\cos\f}\,e^{ i l  \x}
\cdot \frac{\O(k,\3y)}{\m^2+ l ^2} \notag\\
&=\int_0^\8 h\db h\ J_0(h\r) \ir\db l \ 
e^{i l  \x}\cdot \frac{\O(k,\3y)}{\m^2+ l ^2}  \lab{U}
\end{align}
is identical with $B_\o(\3z)$, provided the integration contour is chosen to give the known behavior of $W$. Note that
\begin{align*}
\O(k,\3y)&=\frac1\m\LB \m\cos(\m a)+ l \sin(\m a)\RB\\
&=\frac1{2\m}\LB (\m-i l ) e^{i\m a}+  (\m+i l ) e^{-i\m a}\RB,
\end{align*}
giving 
\begin{align}  \lab{Omega}\bbx{\ 
\frac{\O(k,\3y)}{k^2}
=\frac{e^{i\m a}}{2\m(\m+i l )} +\frac{e^{-i\m a}}{2\m(\m-i l )}\,. \ }
 \end{align}
This  key identity  will reveal  how the  `focusing filter'  $\O$   amplifies forward waves and suppresses backward waves.  Inserted into \eq{U}, it gives
\begin{align}\lab{U2}
U(\3z,\o)\=\int_0^\8 \frac{h \db h}{2\m}\  J_0(h\r) 
\ir\db l \ e^{i l  \x}\,
\LB \frac{e^{i\m a}}{\m+i l } +\frac{e^{-i\m a}}{\m-i l }\RB\!.
\end{align}
The choice of contour  thus amounts to picking a branch of  $\m=\sr{h^2-\o^2}$. Since the denominators $\m\pm i l $ will give residues at 
$l =\pm i\m$, the plane waves $e^{i l  \x}$ will be \sl propagating \rm when $h^2<\o^2$ and \sl evanescent \rm when $h^2>\o^2$.  Define the  branch 
\begin{align*}
\m=
\begin{cases}
-i\sr{\o^2-h^2}\,, &  h^2\le\o^2\\
\sr{h^2-\o^2}\,, &  h^2\ge\o^2.
\end{cases}
\end{align*}
and note that all its values can be shifted to the right half-plane by the infinitesimal translation $\m\to\m+0$. If we use this branch  in \eq{U2},  then by closing the integration contour  in the upper or lower complex half-plane, depending on the behavior of $e^{i l  \x}$, and using Cauchy's theorem, we obtain
\begin{align*}
e^{i\m a} \ir \db l \ \frac{e^{i l  \x}}{\m+i l }
&=\Q(\x)e^{i\m a} e^{-\m \x} =\Q(\x) e^{-\m \z}, \qq \boxed{\  \z=\x-ia\ }\\
e^{-i\m a}  \ir \db l \ \frac{e^{i l  \x}}{\m-i l }
&=\Q(-\x)e^{-i\m a} e^{\m \x}=\Q(-\x) e^{\m \z}.
\end{align*}
This gives $U$ as the even part of a function $U\9+$ defined by
\begin{align}
U(\3z,\o)&=U\9+(\3z,\o)+U\9+(-\3z,\o), \qqq \3z=(\3\r, \z)\notag\\
U\9+(\3z,\o)&=\Q(\x)
\int_0^\8 \frac{h \db h}{2\m} \ J_0(h\r)\, e^{-\m \z}. \lab{U3}
\end{align}
Thus, using the branch $\m$  results in  all waves propagating in the $+\3y$ direction ($\x>0$) being amplified by
 $e^{i\m a}=e^{\sr{\o^2-h^2} a}$ and all those propagating in the $-\3y$ direction ($\x<0$) being suppressed by its reciprocal.   The evanescent waves decay in both directions, as they should. Note that $U(\3z,\o)$ is even in $\3z$, but not in $\3x$ alone.
That is, while it has a preferred direction in \sl real \rm space, it does not have one in \sl complex space. \rm This is obvious since a  pulsed beam in the $-\3y$ direction will have the same behavior in $-\3x$ as the original one has in $\3x$. 

Now recall from the discussion below \eq{rad} that the retarded pulsed beams propagate  in the direction of $\1u\,\3y$, \ie  along $\3y$ if $u>0$ and along $-\3y$ if $u<0$. But the factor $\Q(\o u)$ in the Fourier transform of the Cauchy kernel forces the signs  of $u$ and $\o$ to be identical on the support of $\1W$. Therefore we need the growing exponential $e^{i\m a}$ associated with the $+\x$ direction if $\o>0$ and the $-\x$ direction if $\o<0$, and  the decaying exponential 
$e^{-i\m a}$ associated with the $-\x$ direction if $\o>0$ and the $+\x$ direction if $\o<0$. 

This shows that  \eq{U3} gives the correct value for $\o>0$ but the wrong one for $\o<0$.  The correct branch of $\m$ for $\o<0$ can be obtained by noting that $\2r(\3z^*)=\2r(\3z)^*$, hence
\begin{align*}
B_\o(\3z^*)^*=\frac{e^{-i\o\2r}}{4\p\2r}=B_{-\o}(\3z).
\end{align*}
Therefore we define $U$ for negative frequencies by
\begin{align*}
U(\3z,-\o)&=U(\3z^*, \o)^*=U\9-(\3z,\o)+U\9-(-\3z,\o)\\
U\9-(\3z,\o)&\=U\9+(\3z^*,\o)^*
=\Q(\x) \int_0^\8 \frac{h \db h}{2\m^*} \ J_0(h\r)\, e^{-\m^*  \z},
\end{align*}
where $\m^*$ is the branch
\begin{align*}
\m^*=
\begin{cases}
i\sr{\o^2-h^2},\  & h^2\le\o^2\\
\sr{h^2-\o^2}, &  h^2\ge\o^2.
\end{cases}
\end{align*}
With this, proving \eq{U} reduces to the following.

\begin{thm}
{ \bf (Generalized Weyl formula)}
The time-harmonic complex-source beam $B_\o(\3z)$ has the following  {\bf\sl angular spectrum} representation:
\begin{align}\lab{Weyl}
\frac{e^{i\o\2r}}{4\p\2r}=
\begin{cases}
U\9+(\3z,\o) , &    \o>0,\  \  \x>0\hb{\rm(large right component)}\\
U\9+(-\3z,\o) , &    \o>0,\  \  \x<0\hb{\rm(small left component)} \\
U\9+(\3z^*,\o)^*, &   \o<0,\  \  \x>0 \hb{\rm(small right component)}\\
U\9+(-\3z^*,\o)^* , &    \o<0,\  \  \x<0 \hb{\rm(large left component)}\\ 
\end{cases}
\end{align}
where, for $\3z=(\3\r, \z)=(\3\r, \x-ia)$,  $\o>0$, and $\x>0$,  
\begin{align} \lab{Weyl2}
U\9+(\3z,\o)=\int_\rr2 \frac{\db\3h}{2\m}\ 
e^{i\3h\cdot\3\r- \m\z}
=U\9+\6{\rm prop}(\3z,\o)+U\9+\6{\rm evan}(\3z,\o),
\end{align}
with the propagating and evanescent parts given by
\begin{align}\lab{prop2}
&U\9+\6{\rm prop}(\3z,\o)
=i\!\int_0^\o\sh{-1} \frac{h \db h}{2\sr{\o^2-h^2}} \  
J_0(h\r)\, e^{i \z \sr{\o^2-h^2} }\\
&U\9+\6{\rm evan}(\3z,\o)=\int_\o^\8 \sh{-1}\frac{h \db h}{2\sr{h^2-\o^2}} \ J_0(h\r)\,  e^{- \z\sr{h^2-\o^2} }.\lab{evan}
\end{align}
The components $U\9+(\3z,\o)$ and $U\9+(-\3z,\o)$ in {\rm \eq{Weyl}} are analytic continuations of one another across the plane $\x=0$, with equal boundary values  on that plane for $\r>a$.
The  jump discontinuity, due to the branch cut $\5D$, is imaginary and given by
\begin{align}
J(\3\r,\o)&
\=\lim_{\e\sea 0}\LB B_\o(\3\r, \e-ia)-B_\o(\3\r, -\e-ia)\RB\notag\\
&=i\,\frac{\Q(a-\r)}{2\p\sr{a^2-\r^2}}\,\cosh(\o\sr{a^2-\r^2}), \lab{jump}
\end{align}
with spectral decomposition
\begin{align}\lab{jumpspec}
J(\3\r,\o)=i\int_\rr2 \db \3h\  e^{i\3h\cdot\3\r}\,\frac{\sin(\m a)}{\m}
=i\int_0^\8 h \db h \ J_0(h\r)\,\frac{\sin(\m a)}{\m}\,.
\end{align}

\end{thm}

\bf\sl Remarks \rm 

\bull In the limit $a\to 0$,   \eq{Weyl}   becomes Weyl's  angular-spectrum decomposition of the fundamental solution  of Helmholtz's equation  (see \ci{MW95}, pp.~120--125, where $m=i\m$),
\begin{align*}
\frac{e^{i\o r}}{4\p r}=\int_\rr2\frac{\db\3h}{2\m}\ 
e^{i\3h\cdot\3\r- \m|\x|},\qq r=\sr{\r^2+\x^2},\ \  \o>0.
\end{align*}

\bull  Since $\2r$ depends only on the squares of the components of $\3z$,  it suffices to reverse only the component $\z=\x-ia$ in 
the $\bh y$ direction. However,   the beams are cylindrically symmetric about the $\bh y$ axis,  therefore
\begin{align*}
B_\o(\3\r, - \z)=B_\o(-\3\r, - \z)=B_\o(-\3z)
\end{align*}
and reversing $ \z$ is equivalent to reversing $\3z$.
 
\sv1

\bf\sl Proof. \rm  

Equation \eq{Weyl} is a direct consequence of Formula (26) on page 9 of  \ci{E54}. With
\begin{gather*}
x=h,\ \ b=\o,\ \  y=\r,\  \  \a= \z=\x-ia,\  \  \x>0\\
\sr{y^2+\a^2}=\sr{\r^2+ \z^2}=\2r,
\end{gather*}
it states that the function
\begin{align*}
f\0h=
\begin{cases}
\pm i \sr{h/(\o^2-h^2)}\, e^{\pm i  \z\sr{\o^2-h^2}}, & 0<h<\o\\ &\\
\sr{h/(h^2-\o^2)}\, e^{-  \z\sr{k^2-\o^2}}, & \o<h<\8
\end{cases}
\end{align*}
has \sl  Hankel transform \rm 
\begin{align}\lab{hankel}
\int_0^\8 dh\ f\0h J_0(h\r)\,\sr{h\r}
=\frac{\sr{\r}\,e^{\pm i \o\2r}}{\2r}\,,
\end{align}
which gives \eq{Weyl} for positive $\o$ and $\x$. 
The other cases follow from the symmetries discussed above. 

That $U\9+(\3\r,-\x+ia ,\o)$  is the analytic continuation of 
$U\9+(\3\r, \x-ia,\o)$ to $\x<0$ follows from the known analyticity of $B_\o(\3z)$ outside of the branch cut $\5D$. Recall that
\begin{align*}
\x\to\pm 0\imp \2r=\sr{\r^2+(\x-ia)^2}\to 
\begin{cases}
\mp i\sr{a^2-\r^2}, & 0\le\r\le a\\
\sr{\r^2-a^2}, & \r\ge a,
\end{cases} 
\end{align*}
hence
\begin{align*}
B_\o(\3\r, \x-ia)\to
\begin{cases}
\pm i\, e^{\,\pm\o\sr{a^2-\r^2}}/4\p \sr{a^2-\r^2}, & 0\le\r\le a\\
e^{i\o\sr{\r^2-a^2}}/4\p\sr{\r^2-a^2}, & \r\ge a
\end{cases} 
\end{align*}
and the jump across $\x=0$ is indeed given by \eq{jump}.
By \eq{Weyl2}, the boundary values of the propagating and evanescent parts are
\begin{align*}
&\lim_{\e\sea 0}U\9+\6{\rm prop}(\3\r, \e-ia,\o)
=i\!\int_0^\o\sh{-1} \frac{h \db h}{2\sr{\o^2-h^2}} \  
J_0(h\r)\, e^{a \sr{\o^2-h^2} }=iA\\
&\lim_{\e\sea 0}U\9+\6{\rm prop}(\3\r, ia-\e,\o)
=i\!\int_0^\o\sh{-1} \frac{h \db h}{2\sr{\o^2-h^2}} \  
J_0(h\r)\, e^{-a \sr{\o^2-h^2} }=iA'\\
&\lim_{\e\sea 0}U\9+\6{\rm evan}(\3\r, \e-ia,\o)
=\int_\o^\8 \sh{-1}\frac{h \db h}{2\sr{h^2-\o^2}} \ J_0(h\r)\,  
e^{ia\sr{h^2-\o^2} }=B+iC\\
&\lim_{\e\sea 0}U\9+\6{\rm evan}(\3\r, ia-\e,\o)
=\int_\o^\8 \sh{-1}\frac{h\db h}{2\sr{h^2-\o^2}} \ J_0(h\r)\,  
e^{-ia\sr{h^2-\o^2} }=B-iC
\end{align*}
with $A, A', B$ and $C$ real.  Hence the real part of $B_\o$ is continuous and the jump  across $\x=0$ is
\begin{align*}
&J(\r,\o)=iA-iA'+2iC\\
&\qq=i\int_0^\o \sh{-1}\frac{h\db h}{\sr{\o^2-h^2}} \  
J_0(h\r)\, \sinh(a \sr{\o^2-h^2})
+i\int_\o^\8 \sh{-1}\frac{h\db h}{\sr{h^2-\o^2}}
 \ J_0(h\r)\, \sin(a\sr{h^2-\o^2} )\\
 &\qq=i\int_0^\8 \frac{ h\db h }{\m}\  J_0(h\r)\sin(\m a).  \sh2\qed 
\end{align*} 
\sv1

Now that we have the correct contour, we also know the Fourier transform of the pulsed beam. By \eq{Omega}, 
\begin{align}  \lab{hatW}
\1W(k,y)=\frac{\1g(\o, u)\O(k, \3y)}{\m^2+l^2}
=-\frac{\1g(\o, u)}{2\m}\LB\frac{e^{i\m a}}{\m+0+i l }
+\frac{e^{-i\m a}}{\m+0-i l}\RB,\qq \o>0. 
 \end{align}

\section{Electromagnetic wavelets revisited} 

The scalar complex point sources  will now be used to construct sources for electromagnetic wavelets. The most direct formulation is in terms of  \sl Hertz potentials, \rm which are reviewed briefly. 

Although Hertz potentials have a long history in electrodynamics, they have been sadly ignored in most modern textbooks. Many books that do mention them have only a short section on the subject, usually under various specialized assumptions, and so it is difficult to see their generality because each text gives only a partial picture.  It is not widely known, for example, that the electric and magnetic `Hertz vectors' (which are often introduced alone, with their partner gauged away) transform as a skew-symmetric tensor under the Lorentz group, are compatible with external currents, do not require the Lorenz\footnote{Apparently due to L.V. Lorenz and  not H.A. Lorentz; see  \ci{PR84, B99}.
} 
condition, and have a very large gauge group containing that of the 4-vector potential.  

A comprehensive theory of Hertz potentials and their gauge freedom was developed beautifully by Nisbet \ci{N55, N57} and reformulated in spacetime tensor form by McCrea \ci{M57}. (See also Kannenberg \ci{Kan87}.)  Nisbet's  formulation, translated to the language of differential forms \ci{AMT88, T96}  (which bring substantial simplification and clarity into the picture), was used to construct electromagnetic wavelets in \ci{K02}. We summarize the main results below.  Please refer to that paper for details. 

Hertz potentials form a skew-symmetric tensor like the electromagnetic 
field itself. They are given in a reference frame by the \sl electric and magnetic Hertz vectors \rm $\3Z_e\,, \3Z_m\,,$ which will be used here in the \sl self-dual \rm combination
\begin{align}\lab{Z}
\3Z\0x=\3Z_m\0x-i\3Z_e\0x.
\end{align}
These vectors are generated by electric and magnetic polarization densities $\3P_e\,, \3P_m\,,$ again represented in the self-dual form
\begin{align}\lab{P}
\3P\0x=\3P_m\0x-i\3P_e\0x.
\end{align}
($\3P_m$ is called the \sl magnetization \rm and usually denoted by $\3M$.) The two fields are connected by the wave equation
\begin{align}\lab{ZP}
\Box \3Z\0x=-\3P\0x,\qqq \Box=\D-\pl_t^2.
\end{align}
The electromagnetic field will be presented in the \sl anti\rm-self dual combination
\begin{align}\lab{F}
\3F\0x=\3D\0x+i\3B\0x,
\end{align}
and Maxwell's equations become
\begin{align}\lab{max}
i\pl_t\3F=\curl\3F-i\curl\3P-i\3J,\qqq \div\3F=\r,
\end{align}
where the constitutive relations (in SI units)
\begin{align}\lab{const}
\3D=\3E+\3P_e,\qq \3B=\3H+\3P_m
\end{align}
were taken into account. We will need only polarization sources, and therefore assume vanishing external charge-current density:  $\3J=\30$ and $\r=0$.
Then $\3F$ is  generated by $\3Z$ according to \ci[p.~80]{BW75}
\begin{align}\lab{FZ}
\3F\0x=i\5L\3Z\0x
\end{align}
where $\5L$ is the operator
\begin{align}\lab{Lop}
\5L \3Z=\curl(\curl\3Z)+i\pl_t\curl\3Z.
\end{align}
External charge-current densities can be included in \eq{ZP}  through \sl stream potentials \rm\ci{N55}. 

As in the scalar case, we first construct sourceless EM wavelets.
These will  split into advanced and retarded parts, which are then the vectorial counterparts of the scalar pulsed-beam wavelets. Thus begin with $\3P=\30$, so that
\begin{align}\lab{wave2}
\Box\3Z\0x=\30 \imp \3Z\0x=\int_C d\2k \ e^{ik\cdot x} \3\z \0k.
\end{align}
Since $k\cdot x=\3k\cdot\3x-\o t$,  \eq{FZ} gives
\begin{align}\lab{F2}
\3F\0x&=i\int_C d\2k \ e^{ikx}
\LB -\3k\times(\3k\times  \3\z  ) +i\o \3k\times  \3\z \RB.
\end{align}
To check that this indeed solves Maxwell's equations, note that \eq{max} reduces to
\begin{gather}\lab{max3}
i\pl_t\3F=\curl\3F\
\imp \o \3f=i\3k\times \3f  \hb{or} \4S\0k\,  \3f\0k=  \3f\0k,
\end{gather}
where $\4S\0k: \cc3\to\cc3$ is the $3\times 3$ matrix function on $C$ defined by
\begin{gather}\lab{Sop}
 \4S\0k\,\3v=i\3n\times\3v,\qq \3n\0k\=\3k/\o\,, \qq \3n^2=1\\
\imp\4S^2\3v=\3v-\3n(\3n\cdot\3v) ,\qq \boxed{\ \4S^3=\4S.\ }\lab{S3}
\end{gather}
Equation \eq{F2} therefore  reads
\begin{align}\lab{F3}
\3F\0x=2i\int_C d\2k \ e^{ik\cdot x}\,\o^2\,\4P\0k  \3\z  \0k, \hb{where}
\4P\0k=\frac{1}{2}\lp \4S^2+\4S \rp .
\end{align}
By  \eq{S3}, $\4S\0k$  has the nondegenerate spectrum 
$\{1, 0, -1\}$, and \eq{max3} requires $ \3f $ to have 
eignevalue 1.\footnote{$\4S  \3f = \3f $ means $\3F$ is anti-selfdual, $\4S\3f =-\3f $ means $\3F$ is selfdual,  and  $\4S\3f =\30$ means $\3f $ is longitudinal and $\3F=\30$. 
}
But
\begin{align}\lab{PP}
\4S\4P=\4P=\4P^*=\4P^2,
\end{align}
so $\4P\0k$ is precisely the orthogonal projection to the eigenspace with eignevalue 1.  This shows how Hertz potentials work in Fourier space for free fields.

By \eq{F3}, $\3F$ has coefficient function
\begin{align}\lab{F4}
\3f =2i\o^2\4P\,  \3\z  =\4P\,\3f .
\end{align}
Next,  extend $\3Z\0x$ and $\3F\0x$ to $\5T$  with the analytic-signal transform,
\begin{align}
\bt Z\0z&=\1u\int_C d\2k \  \Q(\o u)\,e^{ik\cdot z}   \3\z  \0k\lab{Z5}\\
\bt F\0z&=\1u\int_C d\2k \  \Q(\o u)\,e^{ik\cdot z} \4P\0k\3f \0k. \lab{F5}
\end{align}
The positive and negative-frequency parts of $\3f $ also have positive
and negative \sl helicities \rm \ci{K94}, so  the restrictions of $\bt F\0z$ to $\5T\6+$ and $\5T\6-$ are positive and negative-helicity solutions. 

To construct the wavelets we need a Hilbert space of solutions.
The inner product  is uniquely determined in Fourier space up to a constant by Lorentz invariance to be
\begin{align*}
\la\3F_1 | \3F_2\ra=\int_C\frac{d\2k}{\o^2}\  \ \3f _1\0k^*\3f _2\0k
=4\int_Cd\2k\ \o^2\,  \3\z  _1^*\4P\,  \3\z  _2\,.
\end{align*}
Denote the Hilbert space of all solutions with finite norm by
\begin{align}\lab{HS}
\5H=\{\3F: \|\3F\|^2=\la\3F|\3F\ra<\8\}.
\end{align}
The wavelets will be \sl dyadics, \rm and to streamline the notation, we 
rewrite the inner product by thinking  of $|\3F_1\ra$ as an (infinite-dimensional) \bf\sl column vector \rm and $\la\3F_1|$ as its adjoint \bf\sl row vector \rm with respect to the above inner product:
\begin{align}\lab{star}
\la\3F_1 |=\3F_1^*:\  \5H\to\4C,&&
|\3F_2\ra=\3F_2: \4C\to\5H, &&
\la\3F_1 | \3F_2\ra=\3F_1^*\3F_2\,,
\end{align}
where $\3F_2: \4C\to\5H$ is the map of scalar multiplication $c\mt c\3F_2$\,. This is simply Dirac notation without bras and kets, called  \sl star notation \rm in \ci{K94}, but somewhat more flexible because now $\3F_1^*$ can be replaced by an \sl operator. \rm

Note that the measure $d\2k/\o^2= \db\3k/2|\3k|^3$ is invariant under  scaling. In fact, an equivalent inner product has been shown to be invariant under the   \sl conformal group \rm $\5C$  of Minkowski space \ci{Gr64}. Therefore the Hilbert space of anti-selfdual solutions
carries a unitary representation of $\5C$.  

Returning to  \eq{F5}, define the matrix-valued function
\begin{align}\lab{Wz2}
\bbh W_z\0k =\1u\, \Q(\o u)\,e^{-ik\cdot z^*}\,\o^2\4P\0k,
\end{align}
represented in spacetime by
\begin{align}\lab{Wz3}
\4W_z(x')=\1u \int_C d\2k\  \Q(\o u)\,e^{ik\cdot(x'- z^*)}\,\o^2\4P\0k
\=\4W(x'-z^*).
\end{align}
This is matrix-valued solution of Maxwell's equations (\ie every column is a solution) designed so that its vector-valued inner product with any solution $\3F\in\5H$ is $\bt F\0z$:
\begin{align}\lab{eval2}
\4W_z^*\,\3F\=\int_C\frac{d\2k}{\o^2}\ \bbh W_z\0k^*\3f \0k=\bt F\0z.
\end{align}
This is a vector form of the \sl evaluation maps \rm  \eq{eval} used to define the scalar relativistic coherent states $e_z$. 

The $\4W_z$s are the the sourceless electromagnetic wavelets.
They  span a \sl reproducing kernel Hilbert space \rm with a matrix kernel  
\begin{align}
\4K(z',z^*)&\=\4W_{z'}^*\4W_z=\int_C \frac{d\2k}{\o^2}\  
\Q(\o u)\,\Q(\o' u)\, e^{ik\cdot(z'-z^*)}\o^4\4P\0k\nt\\
&\=\Q(-y'\cdot y)\4W(z'-z^*),\qqq z', z\in\5T,\lab{RK}
\end{align}
where the factor $\Q(-y'\cdot y)$  enforces the orthogonality of wavelets parameterized by the forward and backward tubes and the holomorphic matrix function
\begin{align}\lab{W}
\4W\0z=\int_C d\2k\  \Q(-k\cdot y)\,e^{ik\cdot z}\,\o^2\4P\0k
=\int_{C\6\pm} d\2k\  e^{ik\cdot z}\,\o^2\4P\0k,\qqq z\in\5T\6\pm
\end{align}
generates the entire wavelet family by  translations: 
\begin{align}\lab{Wz4}
\4W_z(x')=\1u\4W(x'-z^*).
\end{align}

We now compute the `mother wavelet'  $\4W\0z$ explicitly.   Applying it to a vector $\3p\in\cc3$ gives, by the same reasoning as in \eq{F2} and \eq{F3},
\begin{align}\lab{Wp}
2\4W\0z\3p=2\as\o^2\4P\0k\3p=-\5L [G_4\0z\3p],
\end{align}
where, according to \eq{Gz} and \eq{eucl},
\begin{align}\lab{2R}
G_4\0z=-\int_C d\2k\  \Q(-k\cdot y)\,e^{ik\cdot z}
=\frac1{4\p^2z^2}\,,
\qq z\in\5T
\end{align}
is the extension \eq{eucl} of the Euclidean potential $G_4(x\6E)$. Since $\3p\in\cc3$ is arbitrary, \eq{Wp} determines $\4W\0z$.

Thus we have made contact with the scalar theory. We already have a good understanding of $G_4\0z$ and its relation to causality. By \eq{prop},  we have a splitting 
\begin{align}\lab{splitp}
G_4\0z\3p=i\2D\9-\0z\3p-i\2D\9+\0z\3p,\qqq 
\2D\9\pm\0z=\frac 1{8i\p^2\2r(\t\mp\2r)}\,.
\end{align}
It is natural to define retarded and advanced \sl  Hertzian dipole pulsed-beam potentials \rm
\begin{align}\lab{Zpm}\bbx{
\3Z_\3p\9\pm\0z=\2D\9\pm\0z\3p}
\end{align}
whose polarization sources are
\begin{align*}
\3P_\3p\0z=-\Box \3Z_\3p\9\pm\0z
=-\3p\,\Box \2D\9\pm\0z=\3p\,\2\d_{3,1}\0z.
\end{align*}
Thus $\3p$ is interpreted as  a combination of magnetic and electric dipole moments
\begin{align}\lab{pepm}
\3p=\3p_m-i\3p_e
\end{align}
and $\3Z\9\pm_\3p(x-iy)$ are interpreted as emitted by a  \sl dipole disturbance initiated at $iy$ and observed at \rm$x$. By \eq{FZ}, the associated  pulsed-beam fields are
\begin{align}\lab{FZpm}
\3F_\3p\9\pm\0z=i\5L\3Z_\3p\9\pm\0z.
\end{align}
Thus \eq{splitp} induces a  causal splitting of $\4W\0z$,
\begin{align}\lab{Wpm}
\4W\0z\3p=\4W\9+\0z\3p-\4W\9-\0z\3p,
\end{align}
where $\4W\9+\0z$, interpreted as retarded and advanced  \bf\sl electromagnetic pulsed-beam propagators, \rm give $\3F_\3p\9\pm$ directly from $\3p$ by 
\begin{align}\lab{Wpm2}
2\4W\9\pm\0z\3p=\3F_\3p\9\pm\0z.
\end{align}
Since $\2D\9\pm\0z$ is holomorphic in $\5T$  outside the world tube $\tc D$ swept out by the source disk $\5D$,  so are 
$\3Z\9\pm_\3p$, $\3F\9\pm_\3p$ and $\4W\9\pm$. The dipoles are spread over the source disk $\5D$ and modulated in time, as specified by the distribution $\2\d_{3,1}$  in \eq{2d31}.  Note that since $\2\d_{3,1}$ is complex,  the electric and magnetic dipoles become thoroughly mixed while being `translated' from the origin to $iy$. 

The Fourier transforms in $x$ of $\4W\9\pm(x-iy)$ can be computed from those of the scalar sources. For $\o>0$,  \eq{Zpm} and \eq{hatW} with $\1g(\o, u)=\1C(\o, u)$ give
\begin{align}  \lab{hatZp}\bbx{
\bh Z\9+_\3p(k,y)
=\frac{\1C(\o, u)}{2\m}\LB\frac{e^{i\m a}}{\m+0+i l }
+\frac{e^{-i\m a}}{\m+0-i l}\RB\3p. }
\end{align}
The EM wavelets (with $\o>0$) are now obtained in Fourier space by taking the transform of \eq{FZpm} and remembering that $\4S+\4S^2\ne 2\4P$ since we are not on-shell:
\begin{align}\lab{hatWpm}\bbx{
2\wh{{\4W}}(k, y)\3p=
\frac{\1C(\o, u)}{2\m}\LB\frac{e^{i\m a}}{\m+0+i l }
+\frac{e^{-i\m a}}{\m+0-i l}\RB
\LB i\3k\times(\3k\times\3p)+\o\3k\times\3p \RB.}
\end{align}

\bf\sl Remarks \rm

\bull \sl All quantities are real \rm in spite of the the fact that we are dealing with holomorphic functions in complex spacetime! For example,
\begin{align}\lab{Fz}
\3D\9\pm_\3p(x-iy)&=\re \3F\9\pm_\3p(x-iy)\\
\3B\9\pm_\3p(x-iy)&=\im \3F\9\pm_\3p(x-iy)\nt
\end{align}
define a real electromagnetic field in Minkowski space, for any given imaginary source point $iy$.  This justifies our motto, \sl real physics in complex spacetime. \rm 

\bull The holomorphy, rooted in the cone structure of relativistic wave equations as explained in the introduction, serves to organize the equations by pairing dual or `harmonically conjugate' fields. In the case of Maxwell fields, the duality is between magnetic and electric entities. This pairing survives even the introduction of sources in that  \sl local sources introduce only local singularities. \rm  This makes it possible to study the singular sources in terms of the boundary values of the fields, as we have done.

\bull Note that just as holomorphy pairs electric and magnetic \sl dependent \rm variables, it paired positions and momenta as \sl independent \rm  variables in the relativistic coherent-state representations of massive fields. 

\bull It is easily shown that
\begin{align}\lab{conj}
\4P\0k^*=\4P\0k\imp\4W\0z^*=\4W(z^*)\imp \4W_z(x')^*=\4W_{z^*}(x'),
\end{align}
therefore we need only consider $z\in\5T\6+$.

\bull Note that
\begin{align}\lab{homG}
G(z)=s^{-2}\,G(z/s),\qq s\ne 0.
\end{align}
If $s$ is independent of $x$ (but possibly depends on $y$), then
\begin{align}\lab{homW}
\4W\0z=s^{-4}\,\4W(z/s).
\end{align}
Taking $s$ to be the \bf\sl resolution parameter \rm discussed below \eq{Pm},
\begin{align}\lab{res}
s=\sr{-y^2}\=\l>0,
\end{align}
\eq{Wp} and \eq{Wz4} show that
\begin{align}\lab{scale}
\4W_z(x')=\pm\l^{-4}\,\4W\lp \frac{x'-z^*}\l \rp
=\pm\l^{-4}\,\4W\lp \frac{x'-x}\l-i\1y \rp, \qq z\in\5T\6\pm 
\end{align}
where 
\begin{align*}
\1y=\frac y \l\,\in V\6\pm\,,\qq \1y^2=-1.
\end{align*}
Thus all the wavelets are obtained from $\4W(x-iy)$ with $y$ on the hyperboloid $y^2=-1$. Using Lorentz invariance, $y$ can be further restricted to $y=(\30, \pm i)$, which may be further reduced to 
$y=(\30,  i)$  by \eq{conj}. In this way,  the entire family of EM  wavelets is obtained from a single matrix function. Furthermore, the columns and rows of $\4W$ are constrained by \eq{Wz2} since $\4P$ is the projection matrix to a one-dimensional subspace.

\bull On the other hand, because the splitting $G_4=i\2D\9+-i\2D\9-$ depends on the inertial frame where the branch cut is taken (the rest frame of $\5D$), we cannot apply Lorentz covariance to the pulsed beams $\4W\9\pm_z(x')$.
But $\2D\9\pm\0z$ are still \sl positive-homogeneous, \rm 
\begin{align}\lab{homGW}
\2D\9\pm(z)=s^{-2}\,\2D\9\pm(z/s),\qq s >0,
\end{align}
therefore \eq{scale} still holds for $\4W\9\pm_z$.  All these wavelets can therefore be obtained from those on the hyperboloid $y^2=-1$ in $\5T\6+$\,, whose elements $y$ represent the normalized 4-velocity of the source disk.

\bull There exist many equivalent \sl resolutions of
unity in $\5H$ \rm \ci{K94}, obtained by integrating over various parameter sets $\5P\subset\5T$ with appropriate measures $d\m_\5P$:
\begin{align}\lab{RU}
\int_\5P d\m_\5P\0z \4W_z \4W_z^*=I_\5H\,.
\end{align}
This is a `completeness relation' dual to the `(non)-orthogonality' relation \eq{RK}.  One natural subset for a resolution is the \sl Euclidean spacetime \rm  as in \eq{EST}, where all the wavelets are spherical and are  parameterized by their center $\3x$ and scale $u$.

\bull  Each resolution gives a  representation  of EM fields as
superpositions of wavelets,
\begin{align}\lab{RU2}
\3F(x')=\int_\5P  d\m_\5P\0z \4W_z(x')\4W_z^*\3F
=\int_\5P  d\m_\5P\0z \4W_z(x')\bt F\0z,
\end{align}
with the AST $\bt F\0z$ restricted to $\5P $ as the ``wavelet transform.''

\bull Applying a conformal transformation to any resolution of unity gives another one. Since $\5C\app SU(2,2)$ acts on $\5T$ by matrix-valued M\"obius transformations, the new parameter space is simply the transform of $\5P$.

\bull The resolutions of unity considered in \ci{K94} were all continuous, but they can be discretized, provided the sampling is sufficiently fine, to give \sl frames \rm of EM wavelets.

\bull Applying the AST to \eq{RU2} gives
\begin{align}\lab{RU3}
\bt F(z')=\int_\5P  d\m_\5P\0z \ \4W_{z'}^*\4W_z\bt F\0z
=\int_\5P  d\m_\5P\0z\4K(z',z^*)\bt F\0z,
\end{align}
which explains the term ``reproducing kernel.''

\bull Combining \eq{Wpm} and \eq{RU2} gives 
\begin{align}
\3F(x')&=\3F\9+(x')-\3F\9-(x') \nt \\
\3F\9\pm(x')&=\int_\5P  d\m_\5P\0z \4W\9\pm_z(x')\bt F\0z,\lab{RU4}
\end{align}
which is interpreted as a resolution of the sourceless field $\3F$ into retarded and advanced fields generated by complex sources distributed over $\5P$ (more precisely, on disks $\5D_z$ parameterized  by $z\in\5P$).

A great many (most!) aspects of EM wavelets, both  theoretical and practical,  remain unexplored.  As previously mentioned, an exciting possibility is that the pulsed-beam wavelets  $\4W\9+_z$ may be realized by simulating their sources. I hope to report on the continuation of this research in the near future.

\section{Appendix: proofs of the source theorems}

We begin with some preliminaries concerning the oblate spheroidal coordinates associated with the complex distance in $\cc 3$ (for $\cc n$, see \ci{K00}),
\begin{align*}
\2r(\3x-i\3y)=\sr{(\3x-i\3y)^2}=p-iq,\qq p\ge 0.
\end{align*}
Let $\nabla$ be the gradient and $\D$ the Laplacian with respect to $\3x$, for given $\3y\ne 0$. Then
\begin{align*}
&\2r ^2=\3z^2\imp \2r  \,\nabla\2r  =\3z,\qq (\nabla \2r  )^2 =1.
\end{align*}
The unnormalized OS basis $\grad p, \grad q$ is given by
\begin{align*}
\nabla \2r  =\frac{\3z}{\2r}  =\frac{\2r^*\3z}{\2r^*\2r }\imp
\nabla p=\frac{p\,\3x+q\,\3y}{\2r^*\2r}, \qq
\nabla q=\frac{p\,\3y-q\,\3x}{\2r^*\2r}\,.
\end{align*}
Its normalization and orthogonality follow from
\begin{gather*}
\nabla\2r  \cdot\nabla\2r =1 \imp
(\nabla p)^2-(\nabla q)^2=1, \qq \nabla p\cdot \nabla q=0\\
\nabla\2r^*  \cdot\nabla\2r  =\frac{|\3z|^2}{\2r^*\2r}
\imp(\nabla p)^2+(\nabla q)^2=\frac{r^2+a^2}{\2r^*  \2r  },
\end{gather*}
which give
\begin{align*}
(\nabla p)^2=\frac{a^2+p^2}{\2r^*  \2r  }, \qqq
(\nabla q)^2=\frac{a^2-q^2}{\2r^*\2r}.
\end{align*}
Taking the divergence of $ \2r\,\grad\2r  =\3z$ gives \begin{align}\lab{Dr}
\D\2r  =\frac2{\2r},\hb{hence}
\D p=\frac{2p}{\2r^* \2r  }\,,\   \  \D q=-\frac {2q}{\2r^* \2r }\,.
\end{align}
To compute volume integrals in the oblate spheroidal coordinates, 
recall \eq{cyl2} that
\begin{align}
a\x=pq, \qq a^2\r^2=(p^2+a^2)(a^2-q^2).  \lab{cyl3}
\end{align}
Therefore, using differential forms (see \ci{AMT88}, for example), 
\begin{align}\lab{pqVol}
\r d\r\^ d\x&=\frac 12\,d(\r^2)\^d\x
=\frac{1}{2a^3}\,d\lb (p^2+a^2)(a^2-q^2)\rb \^ d(pq)\\
&=a\inv \,(p\,dp-q\,dq)\^(p\,dq+q\,dp)
=a\inv |\2r|^2\,dp \^ dq,
\end{align}
where $dp\^ dq$ denotes the antisymmetric exterior 
product of differential forms. Therefore the volume measure in OS coordinates is
\begin{align}\lab{Vol}
d\3x=a\inv |\2r|^2\,dp\,dq\,d\f.
\end{align}

\sv2

\bf\sl Proof of Theorem \ref{T:shielded}. \rm 

Using the shorthand
\begin{align*}
W_\e\0z=\Q(p-\e a)\,\frac {g(\t-\2r)}{4\p\2r}\=\Q\,\frac g{4\p\2r}\,,\qqq
(\grad p)^2=N,
\end{align*}
we have, taking into account the above relations:
\begin{align*}
-2\p\grad W_\e&=-\d\,\frac g{2\2r}\grad p+\Q\,\frac {g'}{2\2r}\,\grad\2r
+\Q\,\frac {g}{2\2r^2}\,\grad\2r\\
-2\p\D W_\e&=-\d'\,\frac {g}{2\2r}\,N+\d\,\frac {g'}{\2r}\,N
+\d\,\frac {g}{\2r^2}\,N-\d\,\frac {g}{2\2r}\,\D p-\Q\,\frac {g''}{2\2r},
\end{align*}
therefore
\begin{align*}
2\p S_\e=-2\p \square W_\e=-\d'\,\frac {g}{2\2r}\,N+\d\,\frac {g'}{\2r}\,N
+\d\,\frac {g}{\2r^2}\,N-\d\,\frac {g}{2\2r}\,\D p\,.
\end{align*}
Multiplying through by $|\2r|^2$ and letting $\s=p^2+a^2$,
\begin{align*}
2\p |\2r|^2 S_\e=-\d'\, \frac{\s g}{2\2r}
+\d\,\frac{\s g'}{\2r}+\d\,\frac{\s g}{\2r^2}-\d\,\frac{\e g}{\2r}\,.
\end{align*}
Since the only singularities are in $\3x$, no smearing needed in $t$ and $S_\e$ acts on a test function $f(\3x)$ by
\begin{align*}
\la S_\e\,, f\ra&\= \int d\3x\ S_\e(\3x-i\3y, \t) f(\3x)\\
&=\frac1 a\int_0^\8 dp\int_{-a}^a dq \lb-\d'\, \frac{\s g \`f}{2\2r}
+\d\,\frac{\s g'\`f}{\2r}+\d\,\frac{\s g\`f}{\2r^2}-\d\,\frac{\e g\`f}{\2r}\rb,
\end{align*}
where $\`f(p,q)$ is the   mean \eq{mean}  of $f(p,q,\f)$ over $\f$.
Integrating the first term by parts in $p$ and simplifying gives
\begin{align}\lab{Sef3}
\la S_\e\,, f\ra&=\frac{\e^2+a^2}{2a}\int_{-a}^a dq \lb \frac{g'\`f}{\2r}
+\frac{g\`f}{\2r^2}+\frac{g\`f_p}{\2r} \rb_{p=\e}\,.
\end{align}
But $g(\t-\2r)$ is analytic, hence
\begin{align*}
g'(\t-\2r)=-g_p=-ig_q\,.
\end{align*}
Integrating the first term by parts in $q$ gives 
\begin{gather}\lab{Sef2}
\la S_\e\,, f\ra=\frac{\a^*\a}{2a}\lb \frac{g\`f}{i\2r}\rb_{\2r=\a^*}^{\2r=\a}
+\frac{\a^*\a}{a}\int_{-a}^a \frac{dq}{\2r} \  g \`f_\2r\Bigm|_{p=\e}
\end{gather}
where
\begin{align*}
2\`f_\2r=\`f_p+i\`f_q\,.
\end{align*}
This proves \eq{Sef}.  To prove \eq{Se}, simply apply the expression on the right to a test function and integrate using \eq{Vol}.  \sh1\qed

\bf\sl Note: \rm Without the last integration by parts, we would be left with \eq{Sef3}, which defines a perfectly good distribution supported on $E_{\e a}$ \sl without \rm any point sources, given without smearing by
\begin{align}\lab{Se3}
S_\e\0z=\frac{\d(p-\e a)}{4\p}\bigm|\frac\a{\2r}\bigm|^2
\LB\frac{g'}{\2r}+\frac{g}{\2r^2}+\frac{g}{\2r}\,\pl_p\RB\! .
\end{align}
However, when the limit $\e\to 0$ is taken in \eq{Sef3},  the term  $g\`f/\2r^2$ diverges as $q^{-2}$. The integral must therefore be 
\sl regularized \rm by subtracting and adding a Taylor approximation to  the numerator $g\`f$. The resulting limit $\la S, f\ra$ is a \sl nonlocal \rm distribution,  \ie it cannot be written in unsmeared form like  \eq{S}. Therefore, the last integration by parts is needed to obtain  the local expression  for $S\0z$.

\sv2

\bf\sl Proof of Theorem \ref{T:bare}. \rm 

Since the test function  is continuous and $\2r=\pm iq$ denote the same point on $\5D$ (regarded as being in the  upper and lower layer),   we have
\begin{align*}
\`f(-iq)=\`f(iq).
\end{align*} 
Furthermore,  \eq{cyl2} gives
\begin{gather}\lab{plpq}
\pl_p=\frac{p\r}{p^2+a^2}\,\pl_\r+\frac q a\,\pl_\x\,, \qq
\pl_q=-\frac{p^2+a^2}{a^2\r}\,q\pl_\r+\frac p a\,\pl_\x\\
\imp \frac {2a}{\2r}\,\pl_\2r=\frac{a-i\x}{\r}\,\pl_\r+i\pl_\x\, .\notag
\end{gather}
This shows that both partials are antisymmetric in $q$ on $\5D$, with
\begin{align*}
 \`f_p(iq)&=-\`f_p(-iq)=\frac q a\,\`f_\x\\
 \`f_q(iq)&=-\`f_q(-iq)=-\frac q\r\,\`f_\r\\
 \frac {2a}{\2r}\,\`f_\2r&= \frac 1\r\LB a\`f_\r+i\r\`f_\x \RB.
\end{align*}
Inserting this into \eq{Sef2},  taking limit $\e\to 0$, and observing that 
\begin{align*}
\2r(\3z)=\pm ia\imp \3x=\30,\hb{hence} \`f(\pm ia)=f(\30)
\end{align*}
gives
\begin{align*}
\la S\,, f\ra
&=\frac{a}{2}\lb \frac{g(\t+ia)}{a}-\frac{g(\t-ia)}{-a} \rb f(\30)
+a\int_{-a}^a \frac{dq}{\2r} \  g(\t+iq) \`f_\2r(-iq)\\
&=\2g(\t, a)f(\30)+2a\int_0^a \frac{dq}{\2r} \  \2g(\t, q) \`f_\2r(-iq)\\
&=\2g(\t, a)f(\30)+\int_0^a\frac{dq}{\r} \  \2g(\t, q)
\LB a\`f_\r+i\r\`f_\x \RB.
\end{align*}
Changing the integration variable to $\r$ gives
\begin{align*}
\la S\,, f\ra
=\2g(\t, a)f(\30)+\int_0^a\frac{d\r}{q} \  \2g(\t, q)
\LB a\`f_\r+i\r\`f_\x \RB
\end{align*}
where $q=+\sr{a^2-\r^2}$. This proves \eq{Sf}. Again, \eq{S} is proved by applying the right side to a test function and integrating.  \sh1\qed

\sv2

\bf\sl Proof of Corollary \ref{T:cases}. \rm 

For $g_0\0t=\d(t)$, $g\0\t=1/2\p i\t$ and
\begin{align*}
\2g(\t,q)&=\frac 1{4\p i}\lb\frac1{\t+iq}+ \frac1{\t-iq} \rb
=-\frac {i\t}{2\p (\t^2+q^2)}\\
&=\frac {i\t}{2\p z^2}\,,\qqq z\in\tc D.
\end{align*}
Since $W\0z=i\2D\9+\0z$ in this case, that proves  \eq{2d31}. 

For $g_0\0t\=2$,  $g\0\t=\1u=\2g(\t,q)$ by \eq{sign} and 
\begin{align*}
W\0z=\frac{\1u}{4\p\2r}=-\1u\, G_3\0z \imp  S\0z=\1u\,\2\d_3(\3z),
\end{align*}
proving \eq{d3}. Finally, for $g_0\0t=e^{-i\o t}$, 
$g\0\t=\1u \, \Q(\o u)\,e^{-i\o \t}$ and 
\begin{align*}
\2g(\t, q)&=\frac{\1u}2 \, \Q(\o u)\,e^{-i\o \t} (e^{\o q}+e^{-\o q})\\
&=\1u \, \Q(\o u)\,e^{-i\o \t}\,\cosh(\o q).  \sh1\qed
\end{align*}

\sv2

\bf\sl Proof of Theorem \ref{T:hatSe}. \rm 
On $E_{\e a}$ we have
\begin{align*}
\x\0q=\e q,\qq \r\0q=|\h |\,\sr{a^2-q^2}\=|\h|\,\r_0\0q.
\end{align*}
We will need the temporal Fourier transform of $g(\t-\2r)$,
\begin{align*}
\ir dt\ e^{i\o t}g(t-iu-\2r)=\1g(\o, u)\,e^{i\o\2r}
=\1g(\o, u)\,e^{i\e\o a}\,e^{\o q}=A\,e^{\o q}
\end{align*}
since the integration contour can be moved  by $t\to t+\2r$ without crossing the discontinuity of $g(\t-\2r)$ across the real axis. 
(Note that $\4R+\2r=\4R-iq$, so the deformation is bounded uniformly by $q^2\le a^2$.) For the test function in \eq{Sef} choose the plane wave 
\begin{align*}
f(\r, \f, \x)=e^{-i\3k\cdot\3x}=e^{-ih\r\cos\f-il \x}.
\end{align*}
Thus, using \eq{plpq},
\begin{align*}
\`f(\2r)&\=\int_0^{2\p} \db\f\  f(\r,\f, \x)
= J_0(h \r)e^{i l \x}= e^{-i l \x}J_0(h \r)= e^{-i\e l q} J_0(h \r) \\
\frac{2a}{\2r}\,\`f_\2r&=e^{i\e l q}
\LB -\frac{h a-i\e h q}\r\,J_1(h \r)+  l    J_0(h \r) \RB\!.
\end{align*}
Inserting this into  \eq{Sef}  and taking the temporal transform gives
\begin{align*}
\1S_\e(k, y)&=A(\o, u)\,\O_\e(k,\3y)\\
\O_\e(k,\3y)&=\frac{|\h|^2 a}{2i}\LB \frac{e^{\o_\e  a}}{\h a}
-\frac{e^{-\o_\e  a}}{\h^* a} \RB
+\frac{|\h|^2 }2\!\int_{-a}^a dq\ e^{\o_\e  q}
\LB -\frac{h a-i\e h q}\r\,J_1(h\r)+  l   J_0(h\r)  \RB.
\end{align*}
Recalling that $h_\e=|\h|h$ and thus $h\r=h_\e \r_0 $\,, this simplifies to
\begin{gather}\lab{hatSe2}
\O_\e(k,\3y)= \cosh(\o_\e  a)-i\e \sinh(\o_\e  a)-I_1+i\e I_2+\h^*\h I_3
\end{gather}
where
\begin{align*}
I_1&=\frac{h_\e  a}2\int_{-a}^a \frac{dq}{\r_0}\ e^{\o_\e  q}\,
J_1(h_\e  \r_0)
=h_\e  a\int_0^a \frac{dq}{\r_0}\ \cos(i\o_\e  q)\,
J_1(h_\e   \r_0)\\
I_2&=\frac{h_\e  }2\int_{-a}^a \frac{dq}{\r_0}\  qe^{\o_\e  q}\,
J_1( h_\e   \r_0)=a\inv \pl_{\o_\e } I_1\\
I_3&=\frac{  l   }2\int_{-a}^a dq\ e^{\o_\e  q}J_0(h_\e   \r_0)
=  l   \int_0^a dq\  \cos(i\o_\e  q)J_0(h_\e   \r_0).
\end{align*}
The first two integrals  can be evaluated by letting 
\begin{gather*}
q=a\cos\g, \qq  2\y=\m_\e a=a\sr{h_\e^2-\o_\e^2}\,,
\qq  2\c=i\o_\e  a, \qq \y^2-\c^2=h_\e^2 a^2/4\,.
\end{gather*}
Then,  using  \ci[p. 742 (6.688-1)]{GR65},
\begin{align*}
I_1&=h_\e  a\int_0^{\p/2} d\g\ \cos(i\o_\e \cos\g)\,J_1(h_\e  a\sin\g)\\
&=(h_\e  a\p/2)J_{1/2}(\y+\c)J_{1/2}(\y-\c)\\
&=h_\e a \sr{\y^2-\c^2}\,  j_0(\y+\c) j_0(\y-\c)\\
&=\frac{h_\e a}{ \sr{\y^2-\c^2}}\,\sin(\y+\c)\sin(\y-\c)\\
&=2\sin^2\y\cos^2\c-2\cos^2\y\sin^2\c\\
&=2\cos^2\c-2\cos^2\y\\
&=\cos 2\c-\cos 2\y.
\end{align*}
Thus
\begin{align}
I_1&=\cosh(\o_\e  a)-\cos(\m_\e  a)\lab{I12}\\
I_2&=\frac1a \frac{\pl I_1}{\pl\o_\e}
=\sinh(\o_\e  a)-(\o_\e  /\m_\e )\sin(\m_\e  a),  \notag
\end{align}
where we have used
\begin{align*}
\frac{\pl\m_\e}{\pl\o_\e}=-\frac{\o_\e}{\m_\e}\,.
\end{align*}
The third integral is obtained using \ci[p. 737 (6.677-6)]{GR65}:
\begin{align}\lab{I3}
I_3=(  l   /\m_\e )\sin(\m_\e  a).
\end{align}
Inserting  these into \eq{hatSe2} gives  some `miraculous' cancellations
resulting in \eq{hatSe}. The details are given Section  \ref{S:discussion}.  \sh1\qed

\section{Conclusions}

Analytic continuations to complex  time and complex spacetime  abound in physics, although the terminology of `Wick rotations' is, in my opinion, sometimes used too casually, without any mathematical justification or even any \sl basis \rm for justification (`not even wrong'). There have been times while reading papers (or even books) on string theory, for example,  when was unable to tell whether the author was working in a Euclidean or Lorentzian signature.  But even when justified, the extensions are usually regarded as `mathematical methods' without any particular physical significance.  Here is a non-exhaustive list of examples known to me.

\bull  In the correspondence between quantum field theory and statistical mechanics, the imaginary time (more precisely, its period) is related to the reciprocal temperature. But this is regarded as an \sl analogy \rm between the two theories,  albeit a precise and very useful  one. To make it more than analogy one might, for example, interpret the \sl complex \rm time as a combination of evolution and thermal parameters for a system in a \sl local \rm equilibrium state, something like the complex combination of the (also incompatible) position and momentum observables occurring in coherent-state representations. 
(These need not be eigenstates of  a corresponding combinations of operators, as they are in the Bargmann-Segal representation. For example, the relativistic coherent states $e_z$ \eq{ez} do not 
depend on the existence of  `covariant spacetime operators,' which do not in fact exist within the usual framework.

\bull  In Wightman field theory, $n$-point functions are extended to tube domains in their difference variables and powerful methods of complex analysis are used to prove theorems like  PCT and the connection between spin and statistics about the original fields in \sl real \rm spacetime \ci{SW64}.  There is no attempt to interpret the complex coordinates $z=x-iy$, although the interpretation of $y$ as (proportional to) an expected energy-momentum in relativistic coherent states, proved for  free fields in \ci{K77, K78, K87}, extends to general axiomatic fields \ci[Section 5.3]{K90}.

\bull In constructive quantum field theory \ci{GJ87},  the Euclidean region is used to correlate the $n$-point functions of a given theory  by rigorous (Feynman-Kac) path integral methods. Then they are continued back to real spacetime and used to construct interacting fields.  Again there is no attempt to interpret complex spacetime because the quantized field exists only in the Minkowskian region while the random field exists only in the Euclidean region. In between, there are only $n$-point  (Wightman) functions.

\bull There have been various efforts to represent spacetime as a \sl Shilov  boundary \rm of a complex domain (see  \ci{G01} and references therein), but I am not aware of any claiming to do physics directly \sl inside \rm these domains. 

\bull Complex spacetime plays a prominent role in twistor theory \ci{PR86, P87} and the theories of Heaven or H-spaces \ci{HNPT78, BFP80}, but again no direct interpretation is generally given to the 
complex coordinates\footnote{E T Newman, private communication.}

To the best of my knowledge, the only examples (aside from the relativistic coherent states and physical wavelets covered here)  where complex spacetime coordinates are given a direct physical significance have appeared in the works of Newman et al. \ci{N73, NW74}, who have proved the following very intriguing result.

Consider an isolated classical relativistic  system in flat spacetime with positive total `mass' (\ie $m^2\=P_0^2- \3P^2>0$, where $P_\m$ is the total energy-momentum). The total angular momentum splits into orbital and spin parts $\3L+\3s$, and $\3L$ is made to vanish by translating to the center of mass. Similarly, if the system has total charge $e\ne 0$ and a magnetic moment $\3\m$, then its dipole tensor is reduced to $\3\m$ by translating to the center of charge. However: 

\bull With a further  \sl imaginary \rm translation by  $i\3s/mc$, the spin can be made to vanish.  Thus \sl spin may be identified with an imaginary center of mass. \rm

\bull With an imaginary translation by  $i\3\m/e$, the magnetic moment can be made to vanish.  Thus \sl magnetic moment may be identified with an imaginary center of charge. \rm

\bull If the centers of mass and charge coincide, then the spin and magnetic moment can be transformed away simultaneously by an imaginary translation. The necessary and sufficient condition for that is
that  \sl the  gyromagnetic ratio of the system have the Dirac value: \rm
\begin{align*}
\3\m=(e/mc)\3s.
\end{align*}
In the massless case, the world lines with complex center of mass are replaced by a totally null complex plane if the spin (in real Minkowski space) is nonzero.

This idea, although proved in flat spacetime, was inspired by the Kerr-Newman solution to the Einstein equation \ci{N65}, which is the universal model for spinning, charged black holes. It was discovered by performing  a somewhat mysterious complex coordinate transformation on the spherically symmetric solution with mass and charge (Reissner-Nordstr\"om)  which is, roughly, a general-relativistic  version of extending the Newtonian potential from $\rr3$ to $\cc3$, \ie
\begin{align*}
G_3(\3x)=-\frac 1{4\p r}\to G_3(\3z)=-\frac1{4\p\2r}\,.
\end{align*}
The Kerr-Newman solution was soon realized to have  the  Dirac gyromagnetic ratio. Recently,  an old debate was re-ignited with A.~Trautman whether the  Dirac value necessarily depended on the nonlinear character of the equations. Newman settled the question by showing that the Dirac ratio was obtained as well for the linearized solution \ci{N02}. In the related work \ci{K01a}, the charge-current distribution for a (real, static) electromagnetic field defined as in \ci{N73} by a holomorphic Coulomb potential
\begin{align}\lab{hcp}
\3E(\3x-i\3y)-i\3H(\3x-i\3y)=-\grad \,\frac 1{4\p\2r}
=\frac{\3x-i\3y}{\2r^3}
\end{align}
was computed and shown to represent a \sl rigidly spinning disk \rm (the branch cut $\5D$ of $\2r$) with radius $a=|\3y|$ and angular velocity
\begin{align*}
\3\o=c\3y/a^2,
\end{align*}
so that the rim moves at the speed of light. This is consistent with the fact that \eq{hcp} represents the electromagnetic part of the linearized Kerr-Newman black hole.

\section*{Acknowledgements}

I thank Louis Boutet de Monvel for urging me to study hyperfunction theory, and Iwo Bialynicki-Birula, Ted Newman, Ivor Robinson and Andrzej Trautman for friendly discussions and suggestions over the past several years. I also thank David Park for generous and expert help in producing  the plots using his DrawGraphics Mathematica package. Finally, I am grateful to Arje Nachman for his sustained support of my work, most recently through AFOSR Grant  \#F49620-01-1-0271, and to Bill Miceli for support through ONR in the past, when many of the ideas expounded here were just being formed.

\small
\bl {1}

\end{document}